\documentclass[twocolumn]{aastex631}

\newcommand{\msunh}{\>h^{-1}\rm M_\odot}

\newcommand{\mpch}{\>h^{-1}{\rm {Mpc}}}
\newcommand{\kpch}{\>h^{-1}{\rm {kpc}}}

\newcommand{\rmd}{{\rm d}}

\usepackage{color}
\submitjournal{ApJ}

\shorttitle{Halo properties and mass functions}
\shortauthors{Wang et al.}
\graphicspath{{./}{figures/}}
\begin{document}

\title{Halo Properties and Mass Functions of Groups/Clusters from the DESI Legacy Imaging Surveys DR9}

\author{Jiaqi Wang}
\affiliation{Department of Astronomy, School of Physics and Astronomy, and Shanghai Key Laboratory for Particle Physics and Cosmology, Shanghai Jiao Tong University, Shanghai 200240, China}

\author{Xiaohu Yang}
\affiliation{Department of Astronomy, School of Physics and Astronomy, and Shanghai Key Laboratory for Particle Physics and Cosmology, Shanghai Jiao Tong University, Shanghai 200240, China}
\affiliation{Tsung-Dao Lee Institute and Key Laboratory for
    Particle Physics, Astrophysics and Cosmology, Ministry of Education,  Shanghai Jiao Tong University, Shanghai 200240, China}
    
\author{Jun Zhang}
\affiliation{Department of Astronomy, School of Physics and Astronomy, and Shanghai Key Laboratory for Particle Physics and Cosmology, Shanghai Jiao Tong University, Shanghai 200240, China}

\author{Hekun Li}
\affiliation{Department of Astronomy, School of Physics and Astronomy, and Shanghai Key Laboratory for Particle Physics and Cosmology, Shanghai Jiao Tong University, Shanghai 200240, China}

\author{Matthew Fong}
\affiliation{Department of Astronomy, School of Physics and Astronomy, and Shanghai Key Laboratory for Particle Physics and Cosmology, Shanghai Jiao Tong University, Shanghai 200240, China}

\author{Haojie Xu}
\affiliation{Department of Astronomy, School of Physics and Astronomy, and Shanghai Key Laboratory for Particle Physics and Cosmology, Shanghai Jiao Tong University, Shanghai 200240, China}

\author{Min He}
\affiliation{Department of Astronomy, School of Physics and Astronomy, and Shanghai Key Laboratory for Particle Physics and Cosmology, Shanghai Jiao Tong University, Shanghai 200240, China}

\author{Yizhou Gu}
\affiliation{Department of Astronomy, School of Physics and Astronomy, and Shanghai Key Laboratory for Particle Physics and Cosmology, Shanghai Jiao Tong University, Shanghai 200240, China}

\author{Wentao Luo}
\affiliation{CAS Key Laboratory for Research in Galaxies and Cosmology, University of Science and Technology of China, Hefei, Anhui 230026, China}

\author{Fuyu Dong}
\affiliation{School of Physics, Korea Institute for Advanced Study, 85 Heogiro, Dongdaemun-gu, Seoul, 02455, Republic of Korea}

\author{Yirong Wang}
\affiliation{Department of Astronomy, School of Physics and Astronomy, and Shanghai Key Laboratory for Particle Physics and Cosmology, Shanghai Jiao Tong University, Shanghai 200240, China}

\author{Qingyang Li}
\affiliation{Department of Astronomy, School of Physics and Astronomy, and Shanghai Key Laboratory for Particle Physics and Cosmology, Shanghai Jiao Tong University, Shanghai 200240, China}

\author{Antonios Katsianis}
\affiliation{Department of Astronomy, School of Physics and Astronomy, and Shanghai Key Laboratory for Particle Physics and Cosmology, Shanghai Jiao Tong University, Shanghai 200240, China}

\author{Haoran Wang}
\affiliation{Department of Astronomy, School of Physics and Astronomy, and Shanghai Key Laboratory for Particle Physics and Cosmology, Shanghai Jiao Tong University, Shanghai 200240, China}

\author{Zhi Shen}
\affiliation{Department of Astronomy, School of Physics and Astronomy, and Shanghai Key Laboratory for Particle Physics and Cosmology, Shanghai Jiao Tong University, Shanghai 200240, China}

\author{Pedro Alonso}
\affiliation{Department of Astronomy, School of Physics and Astronomy, and Shanghai Key Laboratory for Particle Physics and Cosmology, Shanghai Jiao Tong University, Shanghai 200240, China}

\author{Cong Liu}
\affiliation{Department of Astronomy, School of Physics and Astronomy, and Shanghai Key Laboratory for Particle Physics and Cosmology, Shanghai Jiao Tong University, Shanghai 200240, China}

\author{Yiqi Huang}
\affiliation{Department of Astronomy, School of Physics and Astronomy, and Shanghai Key Laboratory for Particle Physics and Cosmology, Shanghai Jiao Tong University, Shanghai 200240, China}

\author{Zhenjie Liu}
\affiliation{Department of Astronomy, School of Physics and Astronomy, and Shanghai Key Laboratory for Particle Physics and Cosmology, Shanghai Jiao Tong University, Shanghai 200240, China}

\correspondingauthor{Xiaohu Yang, Jun Zhang}
\email{xyang@sjtu.edu.cn, betajzhang@sjtu.edu.cn}



\begin{abstract}
Based on a large group/cluster catalog recently constructed from the DESI Legacy Imaging Surveys DR9 using an extended halo-based group finder, we measure and model the group-galaxy weak lensing signals for groups/clusters in a few redshift bins within redshift range $0.1 \leqslant z<0.6$. Here, the background shear signals are obtained based on the DECaLS survey shape catalog derived with the \textsc{Fourier\_Quad} method. We divide the lens samples into 5 equispaced redshift bins and 7 mass bins, which allow us to probe the redshift and mass dependence of the lensing signals and hence the resulting halo properties. In addition to these sample selections, we have also checked the signals around different group centers, e.g., brightest central galaxy (BCG), luminosity weighted center and number weighted center. We use a lensing model that includes off-centering
to describe the lensing signals we measure for all mass and redshift bins.
The results demonstrate that our model predictions for the halo masses, bias and concentrations are stable and self-consistent among different samples for different group centers. Taking advantage of the very large and complete sample of groups/clusters, as well as the reliable estimation of their halo masses, we provide measurements of the cumulative halo mass functions up to redshift $z=0.6$, with a mass precision at $0.03\sim0.09$ dex.
\end{abstract}




\section{Introduction} 
\label{sec_intro}

In the current scenario of structure formation and evolution, dark matter halos grow hierarchically from small perturbations in the initial density field \citep{1993MNRAS.262..627L}. They are regarded as the building blocks of our Universe. The abundance, structure and spatial distribution of halos as a function of their host halo mass hold important information regarding the cosmological parameters and structure formation theories \citep{1974ApJ...193..437P,1993MNRAS.262.1023W}. 

As dark matter halos are not directly observable, one needs to use fair mass/gravitational potential tracers to infer their mass and structure information. 
In literature, there are various kinds of scaling relations that have been established to infer halo masses in observations, e.g., using the X-ray luminosity of clusters \citep{2009A&A...498..361P,Fujita_2019_XRay},  
the Sunyaev-Zel\'dovich effect of clusters \citep{1972CoASP...4..173S,YangTianyi_2022_SZ}, the velocity dispersion of galaxy groups \citep{2006A&A...456...23B,2013MNRAS.430.2638M,Elahi_2018_velocity_dispersion}, 
satellite kinematics \citep{Bosch2004,ZhaoZhou_2019_sate_dynm,Lange_2019}, 
the galaxy infall kinematics \citep{Zu2013,Zu2014}, and the abundance matching method \citep{Yang2005,Yang_cata_2007,McGaugh_2021_ABmass}, etc.

However, all the above methods to infer halo masses may either rely on a particular cosmology or need to assume the dynamical state of the tracers \citep[][]{Li2021_Qingyang}, which are in general model dependent. A more direct way of detecting the halo mass and moreover the halo structure is using the gravitational lensing signals from large surveys.    
Using weak lensing to explore the halo properties has great advantages as it can directly measure the total mass distribution between the observer and the source \citep{2001PhR...340..291B,2005astro.ph..9252S}.

%Sloan Digital Sky Survey \citep{York2000} initiates the {\it   galaxy-galaxy} lensing analysis since \cite{Fischer2000}.  It is followed by \cite{Sheldon2004}, who chose MaxBCG sample as lenses binned by richness. \cite{Hirata2003} and \cite{Mandelbaum2005,   Mandelbaum2006} not only studied the halo properties of lenses from SDSS main sample, but also analysed sources of systematics caused by PSF, selection effect, noise rectification effect and claimed that the final signal should subtract the signal from random sample. \cite{Simet2016} used redMaPPer as lenses and tested the consistency of the mass-richness relation.  These galaxy-galaxy lensing signals provided us another method to measure the halo mass of lens systems in consideration.

The halo mass function, which describes the number density and evolution of dark matter halos as a function of their mass, is one of the most important cosmological probes.
During the past decades, great efforts were made to measure the halo mass around certain clusters, detected by the SZ effect  \citep{von_der_linden_robust_2014, hoekstra_canadian_2015, penna-lima_calibrating_2017, sereno_psz2lens_2017} or their X-ray luminosity map \citep{Prada2006, Gruen_2014, smith_locuss_2016}, using the weak lensing signals, so that halo mass scaling relations can be obtained and the abundance of the halos (halo mass function) at the high mass end can be measured. However, since the accuracy of weak lensing mass measurement of individual halos is limited by the quality and number density of background source images as well as their redshift distribution estimations, halo masses can only be reliably measured for a small number of the most massive clusters.
On the other hand, through the stacking of weak lensing signals, one can measure the halo properties to much lower mass systems, e.g., from clusters to groups, and to isolated galaxies \citep{Hirata2003, Mandelbaum2005, Mandelbaum2006, Simet2017, Medezinski_2017, Luo2018}. %However, due to the lack of proper estimation of the total number of systems in the Universe, 
However, since it's difficult to quantify the completeness and obtain very accurate measurement of the abundance of the related lens systems,
most of these studies still focus mainly on obtaining a weak lensing mass calibration for various halo mass indicators. 
 
Recently, \citet[hearafter Y21]{Yang2021} extended the halo-based group finder developed in \citet{Yang2005,Yang_cata_2007, Yang2012} so that it can deal with galaxy samples with photometric and spectroscopic redshifts simultaneously. Based on the DESI Legacy Imaging Surveys, they built the largest group catalog to date within redshift range $0.0<z \leqslant 1.0$, and obtained halo mass estimation for each group by applying a total group luminosity ranking method.
Thanks to this large and {\it complete} galaxy group catalog, %\footnote{\adr{{\it Complete} refer to containing more than 80\% of the true halo in the mock catalog.}}
which gives important and reliable abundance information of the groups,
we have the opportunity to obtain direct measurements of the halo mass functions in a relatively wide redshift range by applying the weak lensing stacking technique.

Based on the stacked weak lensing signals, in addition to the halo abundances (mass functions), one can also constrain other halo properties, including the concentrations and biases of halos, as a function of halo mass for different redshift bins. 
The concentration is a key quantity that characterizes the density structure of dark-matter halos, and can be used to trace the formation history of dark matter halos
\citep[e.g.,][]{Wechsler2002,Tasitsiomi2004,Zhao2009,Ludlow13,Du2015,Xu2021}. 
It depends on the halo mass, redshift, and cosmological parameters \citep[e.g.,][]{2000ApJ...535...30J, Prada2012}.
Halo bias $b_{\rm h}$ is defined as the ratio between the %\mf{is it just halo density contrast? or else it only applies to clusters?} 
cluster halo density contrast $\delta_{\rm h}$ and the dark matter density contrast $\delta_{\rm m}$.
It traces the large scale environment of the halos, and is dependent on halo mass, and various secondary properties which were refereed to as the `assembly bias' \citep[e.g.,][]{Gao2005,Gao2007,Xu_semianalytical_2021} and cosmology \citep[e.g.,][]{Jing1998, Sheth2001, Seljak2004, Tinker2010}. 
Given those dependencies, useful cosmological information can be derived using one or a combination of these measurements \citep[see][for a recent attempt]{Ingoglia_2022}.

To accurately measure the weak lensing signals, one needs both high quality
imaging of background galaxies and accurate image processing procedures. Many groups have developed image processing pipelines devoted to improving the accuracy of shape measurement \citep{Kaiser1995, Bertin1996, Maoli2000,  Rhodes2000, vanWaerbeke2001, Bernstein2002,Bridle2002,  Refregier2003, Bacon2003, Hirata2003, Heymans2005, Miller2007,  Kitching2008, Zhang2010,  Zhang_2011, Bernstein2014, Zhang2015, Zhang2016, Luo2017}.  
Among these, the \textsc{Fourier\_Quad} method developed by \citet{Zhang2015} and \citet{Zhang2019} is a particularly efficient one. It uses the image moments in the Fourier Domain to recover the shear signal, with rigorous (model independent) treatments of the point distribution function (PSF) effect, the background and Poisson noise. It runs with a very high speed ($\sim 10^{-3}$ CPU*sec/galaxy) \citep[see][for the performance comparison of a variety of pipelines]{Mandelbaum2015}, quite suitable for large scale weak lensing measurement. Observationally, the DESI Legacy Imaging Surveys \citep{Dey2019} provides the currently largest sky coverage of extragalactic sources with good imaging qualities in three optical bands ($g, r, z$).
One of DESI's subprograms, the Dark Energy Camera Legacy Survey (DECaLS) images the full DESI equatorial footprint. It provides increased depth and excellent seeing, and has more than a hundred million galaxies in 3 bands, thus provides an ideal data set for weak lensing studies.

In this study, we combine the large group and shear catalogs, both constructed from the DESI Legacy Imaging Surveys, to carry out our investigations. 
The main purpose is to obtain a set of reliable measurements of the (cumulative) halo mass functions in a few redshift bins up to redshift $z=0.6$. Meanwhile, we will also provide the observational constrains on the concentration - halo mass and bias - halo mass relations in a few redshift bins. These sets of measurements can be used in further studies to probe the structure formation theory and cosmological parameters. The structure of this paper is organized as follows. We first describe the data sets used in this study, including the lens samples and the shear catalogs in \S\ref{sec_data}. The model to describe the lensing signals is presented in \S\ref{sec_esd}. The general halo properties obtained from fitting the lensing signals are given in \S\ref{sec_properties}. We provide our estimation of the  cumulative halo mass functions in \S\ref{sec:ahmf}. Finally we summarize our results in \S\ref{conclsion}.
Throughout this paper, we assume a $\Lambda$CDM cosmology with parameter values from the latest Planck Collaboration analysis \citep[][hereafter Planck18]{2020A&A...641A...6P}:
$\Omega_{\rm m} = 0.315$, $\Omega_{\Lambda} = 0.685$, $\sigma_8 = 0.811$,
$n_{\rm s} = 0.965$ and   $h=0.674$.
All distances are in co-moving units of $\mpch$. The unit of projected surface mass density $\Sigma$ and density contrast $\Delta \Sigma$ is $h {\rm M_\odot}/{\rm pc}^2$. In this work, halos are defined to have 180 times the mean background density $M_{\rm 180m}$.

\section{observational data}
\label{sec_data}

In this section we describe the construction of our lens samples and the source shear catalogs. 

\subsection{Lens samples}

We use the group catalog recently constructed by Y21 from the DESI legacy imaging surveys (now updated to DR9) using an adaptive halo-based group finder for galaxies with either photometric or spectroscopic redshifts\footnote{see \url{https://gax.sjtu.edu.cn/data/DESI.html}}.
% \adb{Note that there are quite a number of masses will be used in our presentation. In order to avoid confusion, we list in Table~\ref{tab:name} the various masses that are used in this paper along with their definitions.}

\begin{figure}
%    \hspace{2cm}
    \centering
    \includegraphics[width=0.45\textwidth, height=0.45\textwidth]{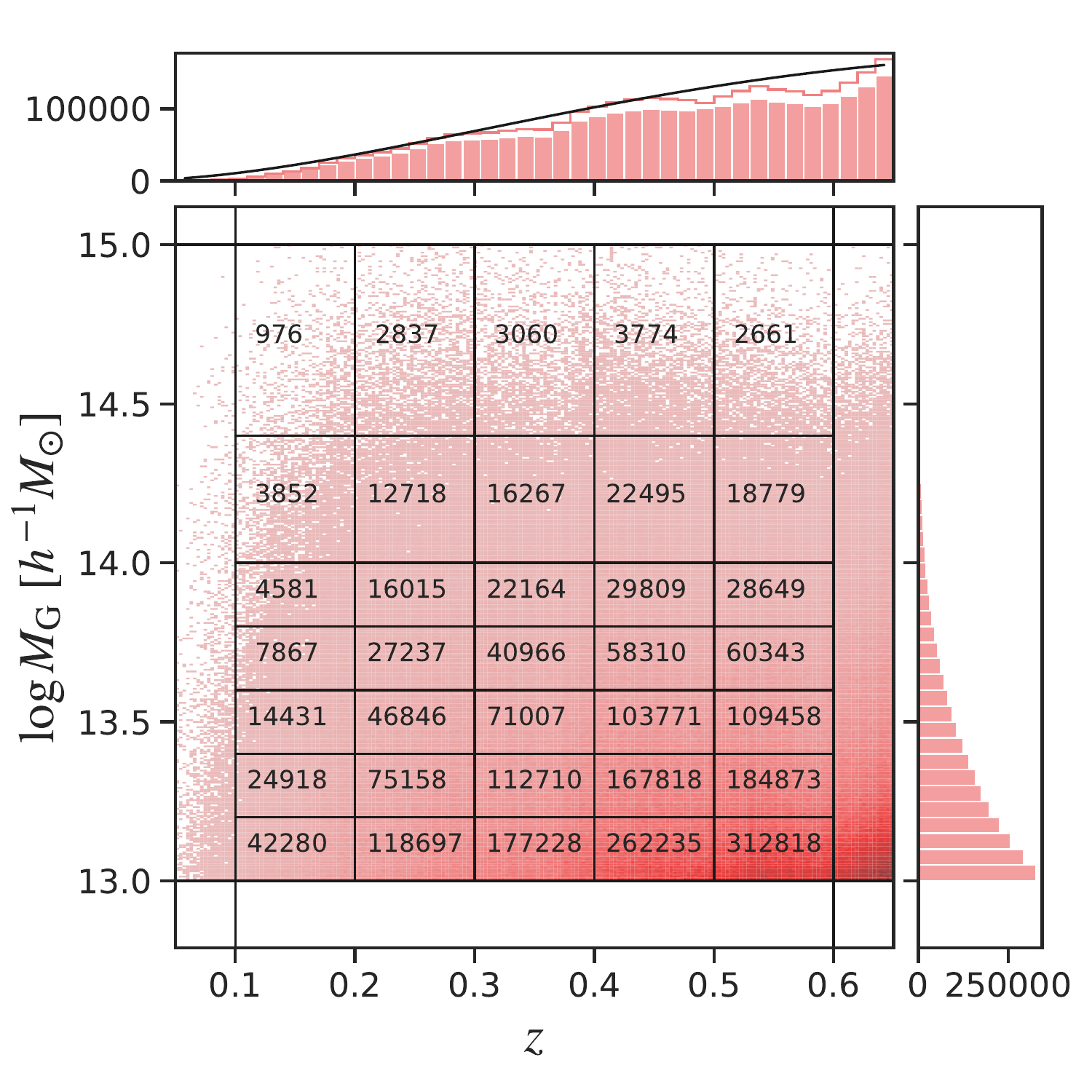}
    \caption{The distribution of group/cluster sample in the redshift and mass proxy space. The $x$-axis is the redshift and the $y$-axis is the group mass. The black solid lines label out thresholds of our redshift and mass bins. The numbers in each block are the number of groups. The top and right panels show the number distributions of groups using 1D shaded histograms according to the redshift and group mass, respectively. In the top panel, we also provide the number distribution of groups after completeness correction (see Eq. \ref{eq:completeness}) using a red histogram and the theoretical prediction of the number distribution of halos with mass larger than $10^{13} \msunh$ under Planck18 cosmology using a black line \citep{Tinker2008}.}
    \label{fig:halo}
\end{figure}

As demonstrated in Y21, using mock galaxy redshift surveys, the group finder can reliably detect most of the groups, especially with mass $\ga 10^{13.0}\msunh$, where the group {\it purity} is larger than 90\%. 
That means more than 90\% of the detected groups are the original true groups in simulations. The halo mass assigned to each group (hereafter, the group mass, $M_{\rm G}$). %, which is defined as $M_{\rm 180m}$)}%\mf{maybe include in the parenthesis: "..., where the group mass is "} 
has an uncertainty of about 0.2 dex at the high mass end ($\ga 10^{14.0}\msunh$), increasing to 0.4 dex at the mass ($\sim 10^{12.3}\msunh$) and then decreasing to 0.3 dex at low mass end ($\sim 10^{11.6}\msunh$). 
%\adr{the standard deviation of the difference between luminosity matched log mass and the true halo mass from the mock catalog}
As the main purpose of this study is to obtain an accurate measure of the halo mass function, where the group {\it purity} is most important, we select
groups/clusters with mass $M_{\rm{G}}>10^{13}\msunh$. %(group mass defined as $M_{\rm 180m}$) \mf{if you defined the group mass above then you won't need it here. but giving all the group mass definitions when it is first mentioned makes it easier to find for the reader}
Note that, as the group masses in the catalog are obtained via a halo abundance matching method, $M_{\rm G}$ in general can be regarded as a measure for the rank-order in the total group luminosity. Here we use group mass rather than the total group luminosity as our rank-order group sample selection criteria, as the groups are extracted from a flux-limited sample and the total group luminosity may suffer from different incompleteness cuts. 
In addition, in order to assure that groups have sufficient richness, 
we only use groups in the redshift ranging from $0.1 \leqslant z<0.6$. 
%\mf{I didn't know about this! I probably should have done the same for my previous paper. you don't have to answer this in this paper, but if you can answer this for me when you have time: What is the reason for this? Does this statement only apply for masses $\log M_{rm G} > 13$?}
We divide all the groups/clusters with the above criteria into 5 redshift bins and 7 mass bins, which results in a total of 35 lens samples. The selection criteria and the number of groups in these lens samples are illustrated in Fig. \ref{fig:halo}. The redshift and mass bins are divided by the solid lines in the figure, and the number of groups in each bin is given in each block. Overall, the vast majority of groups in our 35 lens samples contain at least two members, only a very small fraction of groups at high redshift and low mass bins contain only one member galaxy. 

Once these lens samples are selected, we need to determine the choice of group centers for stacking the lensing signals. 
As pointed out in \citet{Luo2018}, the brightest central galaxy (BCG) is a better tracer of the halo center when compared to the luminosity weighted center (LWC) and the number weighted center (NWC).  
%Note on the other hand, since we are using the photometric redshifts in determining the group memberships, which may suffer from projection effect and the luminosity error induced by the photo-z error. Here we also consider the second brightest galaxy (SBG) as the indicator of the group center. 
However, as we will outline later in Section \ref{sec:model}, since we have taken potential off-centering effect into account in our lensing models, 
in general, we can use the extracted halo information from different sets of halo center indicators for cross checks. 
Thus, we decide to measure the group-galaxy lensing signals around all these three types of group/cluster center indicators. We take the BCG as our fiducial case, and the other two as reference cases.

%We further separate our sample by two second properties. One is the magnitude gap, which gives us implication on galaxy assembly history. Here we use the magnitude gap for each group in z-band. Also, in this paper we use a new property, the distance between the brightest galaxy and the luminosity weighted center, to 

\subsection{The shear catalog}

Our shear catalog is constructed using the imaging data of the DECaLS DR8~\citep{Dey2019}. The survey uses the Dark Energy Camera (DECam) installed on the Blanco 4m telescope. 
It is designed to target sources for the DESI program in the North Galactic Cap region at Dec$ \leqslant 32^{\circ}$ and the South Galactic Cap region at Dec$ \geqslant -34^{\circ}$. 
The images are taken in $g,r,z$ three bands. The total sky coverage by the images is more than 10000 deg$^2$. Our source galaxy catalog is taken from~\cite{Zou2019}, with photometric redshifts of galaxies updated using those obtained by \citet{zhourongpu2021} with a machine learning algorithm, with a typical redshift error $z_{err} / (1+z) = 0.02\sim 0.03$ in redshift range $0.4\leqslant z < 0.9$ (see Fig. 4 in \citet{zhourongpu2021}). 

Shear catalogs are constructed from the DECaLS DR8 using the \textsc{Fourier\_Quad} (FQ) method ~\citep{Zhang2008,ZhangJ_2022,Mandelbaum2015}. FQ is a moment-based method. Its shear estimators are defined with the multipole moments of the 2D galaxy power spectrum, which can correct for the effect of Point Spread Function model-independently. The effects due to the background noise and the Poisson noise can also be removed rigorously in statistics. The method has been tested under general observing conditions to a very low Signal-to-Noise Ratio (SNR) of the source image (SNR $< 10$)~\citep{Zhang2015}. The full image processing pipeline based on the FQ method has been developed and applied on the CFHTLenS~\citep{Heymans_2012,Erben_2013} and the DECaLS data. In both cases, the resulting shear catalogs can successfully recover the small optical field distortion signals ($\sim 1-5\times 10^{-3}$) that are originally derived from the astrometric calibrations, demonstrating its robustness in practice~\citep{Zhang2019,Wang_2021}. In the development of the FQ pipeline, we have carefully studied the source selection effect due to the inevitable incompleteness of sources at the faint end ~\citep{Li_2021}, and also included an algorithm to avoid the systematic errors (additive) due to the presence of the geometric boundaries \citep[CCD edges, bad columns, etc.][]{Wang_2021}.

An interesting feature of FQ is that it tolerates the existence of very poorly resolved images or even point sources in the galaxy samples. This property is particularly important for our processing of the DECaLS data, as the typically full width at half maximum (FWHM) of the PSF is about 1.4-1.5 arcsec, much larger than those of other weak lensing surveys. Overall, our FQ image processing pipeline for the DECaLS data is very similar to that of~\cite{Zhang2019}, except for some details regarding the PSF reconstruction \citep[see][for the details of the construction of DECaLS DR8 shear catalog] {ZhangJ_2022}.

Since FQ works on individual exposures, in our DECaLS shear catalog, images of the same galaxy on different exposures contribute independent shear estimators. As the $g$-band image quality is not as good as the $r$ and $z$ -bands, we decide to use the $r$ and $z$ band shear catalogs for our calculations in this paper. Note that the multiple (multi-band) shear estimators from each galaxy are used together in our measurement of this work. On average, the catalogs contain results from more than 10 source images per square arcmin, with a median redshift at around 0.55.

%\begin{turnpage}
\begin{figure*}
%    \hspace{2cm}
    \centering
    \includegraphics[width=0.93\textwidth]{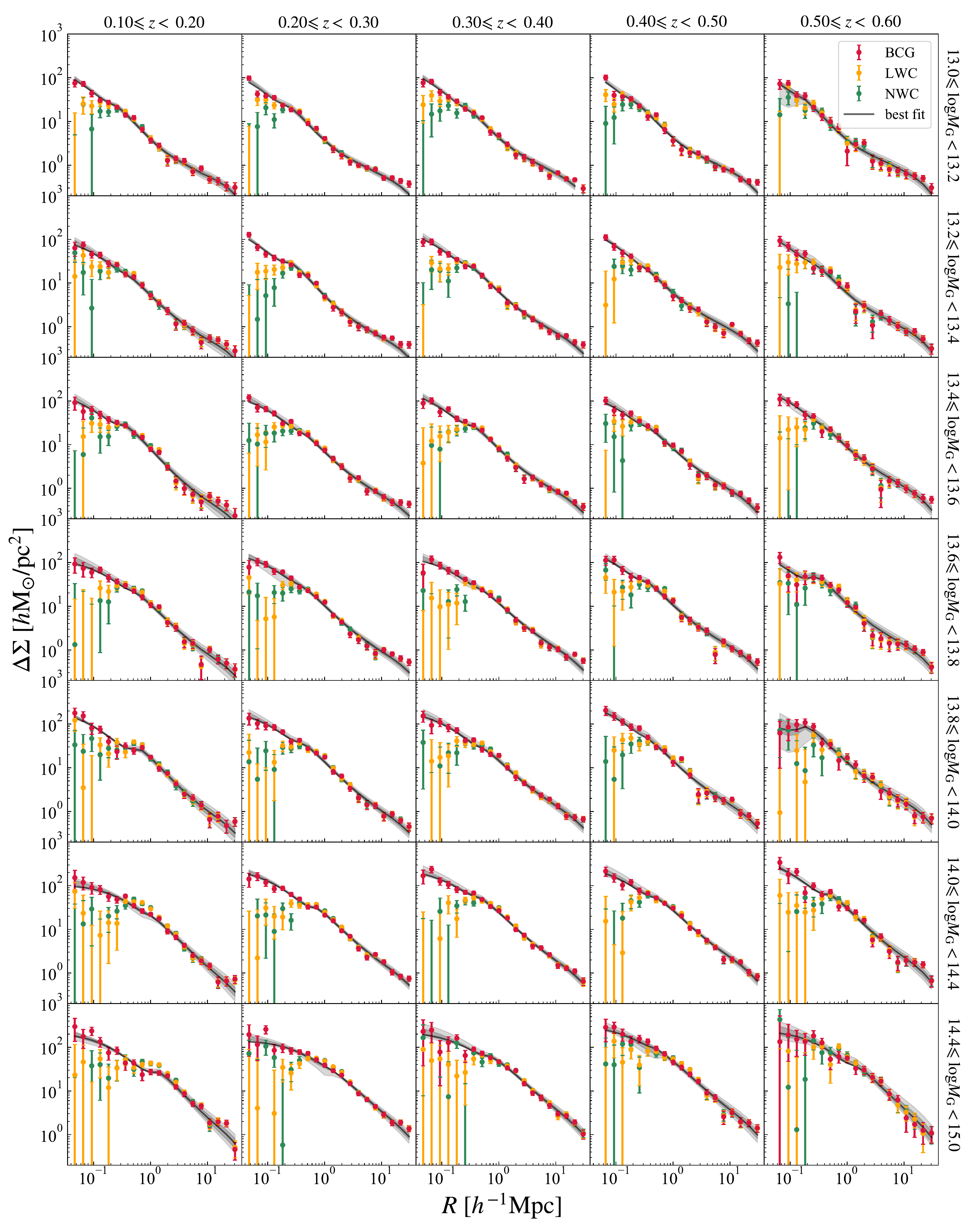} 
    \caption{The excess surface density of DESI Legacy Imaging Surveys DR9 lens samples in different redshift and mass bins as indicated in different columns and rows. In each panel, we compare results for three different group center indicators. The red points represent results for the BCG group center, orange for the LWC and green for the NWC. We can see that BCG performs the best as a center indicator. The solid line in each panel shows the best-fit model (Eq. \ref{eq:esdmodel}) of the ESD for BCG centroid.  The dark and light grey shaded regions represent the 68 and 95 confidence regions of the ESD fitting results respectively.} 
    \label{fig:esds}
\end{figure*}
%\end{turnpage}

\section{ESD Measurement and modeling} \label{sec_esd}

In this section, we provide our measurement and modeling of the excess surface densities (ESDs) around our lens samples. 

\subsection{ESD Measurement} \label{sec:measure}

The weak lensing signal around foreground galaxy groups/clusters is related to the foreground density profile as: 
\begin{equation}\label{eq:ESD}
\Delta \Sigma(r) = \gamma_{\rm t}(r) \, \Sigma_{\rm c}\,, 
\end{equation}
in which $\Delta \Sigma(r)$ is the excess surface density in comoving units, $\gamma_{\rm t}(r)$ is the tangential shear, and $\Sigma_{\rm c}$ is the critical surface density defined as: 
\begin{equation}
\Sigma_{\rm c} = \frac{c^2}{4\pi G} \frac{D_{\rm s}}{D_{\rm l} D_{\rm ls}(1+z_{\rm l})^2}\,,
\end{equation}
where $D_{\rm s}, D_{\rm l}$ and $D_{\rm ls}$ are the angular diameter distances of the source, the lens, and between the source and the lens, $c$ is the light speed and $G$ is the gravitational constant. %\mf{define all parameters here, including c and G} 
$z_{\rm l}$ refers to the redshift of the lens, and the factor $(1+z_{\rm l})^2$ is included to account for the conversion from the physical units to the comoving ones.

In our measurement, the ESD around different lens are stacked using the PDF-symmetrization (PDF\_SYM) method within each sample.
To approach the lower bound of the statistical uncertainty, i.e. the Cramer-Rao Bound, \cite{Zhang2017} (Z17 hereafter) proposed the PDF\_SYM method, which can maximally utilize the ensemble information of the shear estimators based on the distribution function.
Instead of taking the weighted average of the shear estimators as what is usually done, the idea of Z17 is to find the shear signal that can best symmetrize the PDF of the corresponding shear estimators. For example, for an underlying shear signal $g_1$, its best estimate $\hat{g}_1$ can be found by symmetrizing the PDF of the quantity $G_1-\hat{g}_1(N+U)$, in which $G_1$ is the shear estimator along the specified direction, $N$ is the normalization factor, and $U$ is a spin-4 quantity defined in Z17 to account for the parity symmetry. Note that the value of $U$ is also dependent on the direction of the coordinates.

For the ESD measurement, the operation is quite straightforward. For a given foreground lens sample at $z_l$, instead of shear along a specified direction $\hat{g}_1$, we directly look for the value of the excess surface density $\widehat{\Delta \Sigma}$
that can best symmetrize the PDF of the following quantity:
\begin{equation}
    G_{\rm t}(z_{\rm s}) - \frac{\widehat{\Delta \Sigma}}{\Sigma_c(z_l, z_s)}\cdot (N +U_{\rm t})(z_{\rm s})\,,
    \label{eq:PDFSymm1}
\end{equation}
where $z_s$ stands for the source redshift, while $G_{\rm t}$ and $U_{\rm t}$ are all defined in the tangential direction.  The PDF is made of the quantities defined in Eq. \ref{eq:PDFSymm1} from all the background galaxies in a given radius range of interest. In doing so, we properly take into account the contribution from different background redshifts. Simulations regarding the verifications of the PDF\_SYM method on the measurement of the excess surface density will be reported in a separate work (Li et al., in preparation), in which we also discuss about the effects related to photo-z errors.  %\mf{just in case: check all in prep works to see if they have been published or not}

In this work, to avoid the dilution effect by foreground galaxies, we only use sources with redshift $z_{\rm s} > z_{\rm l} + \Delta z$, with $\Delta z = 0.2$. The photo-z errors of the background sources cause the uncertainty of $\Sigma_{\rm c}$, which in turn leads to bias in the ESD signal, making it slightly higher/lower in the lowest/highest redshift bin of our lens sample. This bias, which ranges from almost zero to about 2\%, are corrected using the redshift PDF (fully discussed in Li et al., in preparation). The covariance matrix in this work is computed using 400 jackknife samples. 

%Our discussion of the potential systematic uncertainties due to, e.g., photo-z biases, dilution by member galaxies, magnification bias, etc. are given in a separate work .
%in which $z_{\rm l}$ and $z_{\rm s}$ are the redshifts of the lens and source respectively,  $G_{\rm t}$. Equation~\ref{eq:PDFSymm1} stems from the well known relation between the excess surface density of the lens and the tangential shear signal of the background: $\Delta \Sigma(r) = \gamma_{\rm t}(r) \, \Sigma_{\rm c}$. Here we assume the cosmology given in Section~\ref{sec_intro}. 

Using the above method, we obtain the ESDs around $5 \times 7$ lens samples with three types of group center indicators. The dots with error bars shown in Fig. \ref{fig:esds} are the measured ESDs, where different colors correspond to different group center indicators. Considering the signal-to-noise ratios of the ESD measurements, we provide one less data point on the smallest scale for lens samples in the two high redshift bins than the three low redshift bins. 
In these lens samples, the ESD amplitudes increase more quickly for samples with higher group mass than for samples with higher redshift.
There are no significant differences for ESDs around different centers at large scales. At small scales ($\leqslant 0.5\mpch$), ESDs of samples using BCG as center indicators keep rising with decreasing central distance, while other two central indicators tend to decrease.
The signal-to-noise ratio is low at small scales, especially for the samples using LWC or NWC centroid, due to the decreasing lens-background pairs and the possible large offsets.

\subsection{ESD Modeling} \label{sec:model}

Here we describe the ESD model we use to extract the halo properties.
Theoretically, the ESD around a lens sample is related to the line-of-sight projection of the group-matter cross correlation function,
\begin{equation}
\xi_{\rm gm}(r)=\langle\delta_{\rm g}({\bf x})\delta_{m}({\bf x}+{\bf r})\rangle\,,
\end{equation}
so that \citep[e.g.]{Yang2006,Bosch2013},
\begin{equation}
\label{eq:sigatr}
\Sigma(R) = 2 \overline{\rho} \int_{R}^{\infty} [1+\xi_{\rm gm}(r)] 
{r \, \rmd r \over \sqrt{r^2 - R^2}}\,,
\end{equation}
and
\begin{equation}
\label{eq:siginr}
\Sigma(\leqslant R) = \frac{4\overline{\rho}}{R^2} \int_0^R y\,\,dy\,
 \int_{y}^{\infty} [1+\xi_{\rm gm}(r)] {r \, \rmd r \over \sqrt{r^2 - y^2}}\,,
\end{equation}
where $\overline{\rho}$ is the average background matter density of the
Universe. In general, from the weak lensing shear catalogs, we are measuring the ESDs around the lens systems, i.e., $\Delta\Sigma(R) =\Sigma(\leqslant R)-\Sigma(R)$. Following \citet[][hereafter L18]{Luo2018}, our ESD model contains
three components:
\begin{equation}
  \Delta\Sigma(R)=\Delta\Sigma_{*}(R)+\Delta\Sigma_{\rm 1h}(R)+\Delta\Sigma_{\rm 2h}(R)\,,
  \label{eq:esdmodel}
\end{equation}
where they correspond to, from small to large scales, the contributions of (1) the stellar mass of the lens galaxy $\Delta\Sigma_{*}(R)$, if BCG is used to indicate the group center; (2) the 1-halo term contribution from the host halo mass taking into account the off-centering effect $\Delta\Sigma_{\rm 1h}(R)$;  and (3) the projected two halo term $\Delta\Sigma_{\rm 2h}(R)$, respectively.
%The stellar component is $\frac{M_{\star}}{\pi R^{2}}$, where 

As demonstrated in L18, stellar
contribution is significant at scales smaller than $50\kpch$, then on  $50\kpch\sim 5\mpch$ scales the host halo contributes most of the ESDs.  The 2-halo term contribution is then important at scales larger than a few virial radii. As we have obtained the ESD measurements for our lens samples over much larger scales, we can obtain both the
1-halo and the 2-halo term properties of dark matter halos in this study.
Below we present each term of our ESD model in Eq. \ref{eq:esdmodel}. Note that only small changes are made with respect to the one provided in L18.%, readers who are familiar with this set of formula can go directly to Section \ref{sec_properties}.

The first term in Eq.\ref{eq:esdmodel} is contributed by the stellar mass of the lens galaxy. As pointed out in
\cite{Johnston2007} and \cite{George2012}, the stellar mass component
can be treated as a point mass, and the related ESD can be modelled as,
\begin{equation}
\Delta\Sigma_{*}(R)=\frac{M_*}{\pi R^2}\,,
\end{equation}
where $M_*$ is the stellar mass of candidate BCGs in
consideration, obtained using the $K$-correction software of ~\citet[][v4\_3]{Blanton2007} by applying to the BCG 5 band apparent magnitudes. 
%We directly use the average stellar mass of BCGs in our lens samples for our BCG  group center indicator, and zero for the LWC and NWC group center indicators.
We only model the contribution of the stellar component (using eq. 7) for the BCG group center indicator, using the average stellar mass of all BCGs in the stack.

The second term in Eq.\ref{eq:esdmodel} is the 1-halo term contribution. Here we divide the ESDs into two components, one for the halo center and the other for an off-centering situation,
\begin{equation}
\Delta\Sigma_{\rm 1h}(R)=f_{\rm cen}\Delta\Sigma_{\rm NFW}(R)+(1-f_{\rm cen})\Delta\Sigma_{\rm off}(R)\,,
\end{equation}
where $f_{\rm cen}$ is the fraction of groups that do not suffer from off-centering effect, $\Delta\Sigma_{\rm NFW}(R)$ is ESD of the halo assuming an NFW profile (see Eq. \ref{eq:SigmaNFW}).
%Here we fix the concentration\footnote{As subhalos are on average relatively low mass ones and may be affected by  stripping effect, their concentration is thus set to relativel y large values.  } as $c=15$ and treat $M_{\rm sub}$ as our first free parameter in our modelling. We require that subhalo mass to be at least a factor of 3 smaller than the host halo mass, $\log M_{\rm h} - \log M_{\rm sub}\ge 0.5$.
For the off-centering halo contribution, $\Delta\Sigma_{\rm off}(R)$, according to \citet{Yang2006}, the projected surface density for a galaxy with a projected off-center distance $R_{\rm off}$ from the NFW halo center, can be described by $\Sigma_{\rm off}(R|M_{\rm h},c,R_{\rm off})$ (Eq. \ref{eq:Sigmaoff}). Here, $M_{\rm h}$ and $c$ are the mass and concentration obtained by weak lensing that correspond to $M_{\rm 180m}$.%Our definition of the halo mass $M_{\rm h}$ and the concentration $c$ are the same as the group mass. \mf{if $M_{\rm h}$ is the same as the group mass, then why not use $M_{\rm G}$ here? Also you don't want to say our definition of $M_h$ and $c$ are the same as $M_g$, since you're saying two things are the same as one. do you mean the mass definition? like $M_{\Delta}$ and $c_{\Delta}$ are such that $\Delta = 180m$? I am not sure if this is what you mean, but I think this is the weak lensing mass and concentration you are fitting for that corresponds to $M_{\rm 180m}$. So instead you can say something like "Here, $M_{\rm h}$ and $c$ are the mass and concentration obtained by weak lensing that correspond to $M_{\rm 180m}$." You can keep it just $c$ but I personally think it is more consistent for the reader to change all your $c$ to $c_{\rm h}$. This also avoids confusion when you define/use $c$ in the critical surface density.}
Here, we adopt the model proposed by \citet{Johnston2007} to take into account the off-centering effect,
\begin{equation}
P(R_{\rm off})=\frac{R_{\rm off}}{R_{\rm sig}^2}\exp(-0.5(R_{\rm
  off}/R_{\rm sig})^2)\,,
\end{equation}
where $R_{\rm sig}$ is the dispersion of $P(R_{\rm off})$.
The resulting projected density
profile is the convolution between the $P(R_{\rm off})$ and the
$\Sigma_{\rm off}(R|M_{\rm h}, c, R_{\rm off})$,
\begin{eqnarray}\label{eq:Sigmahost}
&&\Sigma_{\rm off}(R|M_{\rm h}, c,R_{\rm sig}) = \nonumber \\
&&\int dR_{\rm off}P(R_{\rm off})\Sigma_{\rm off}(R|M_{\rm h}, c,R_{\rm off})\,.
\end{eqnarray}
%In total, we have five free parameters regarding the host halo properties, $M_h$, $c$, $f_{\rm cen}$ and $R_{\rm sig}$, in our ESD modeling. 
%In addition to the off-centering effect, here we also take into account the scatter of the host halo mass by using a log-nomal distribution of halo mass with a fixed scatter $\sigma_h(\log M_h)=0.3$. As pointed out in L18 and \citet{Dong2019}, if the scatter of the host halo mass is not taken into account, this will induce the Eddington bias in halo mass estimation.  

Finally, the last term in Eq. \ref{eq:esdmodel} is contributed by the 2-halo term. 
The contribution from the projected mass density field outside the halo, $\Delta\Sigma_{\rm 2h}(R)$, can be calculated from the halo-mass cross correlation functions using Eqs. \ref{eq:sigatr} and \ref{eq:siginr}. 
Here we adopt the model proposed in \citet{Bosch2013} to describe the 2-halo term cross correlation function,
\begin{equation}
    \xi_{\rm gm,2h}(r)=b_{\rm h}\,\zeta(r,z)\, f_{\rm exc}(r)\,\xi_{\rm mm}(r)\,,
\end{equation}
where $\xi_{\rm mm}(r)$ is the auto correlation function of dark matter calculated from the nonlinear power spectrum \citep[e.g.]{Smith2003}, $f_{\rm exc}(r)$ is to characterize the halo exclusion effect (see Eq. \ref{eq:exclu}), $\zeta(r,z)$ is a function used to take into account the radial dependence of the halo bias (see Eq. \ref{eq:zeta}), and $b_{\rm h}$ is the last free parameter in our ESD modeling describing the bias of the dark matter halos. 

In total, we have five free parameters in our ESD model, $M_{\rm h}, c, f_{\rm cen}, R_{\rm sig}$ and $b_{\rm h}$ for our ESD measurements for all  group centers. Below we outline the processes we use to constrain the model parameters.

\begin{figure*}
    \centering
    \includegraphics[width=0.7\textwidth]{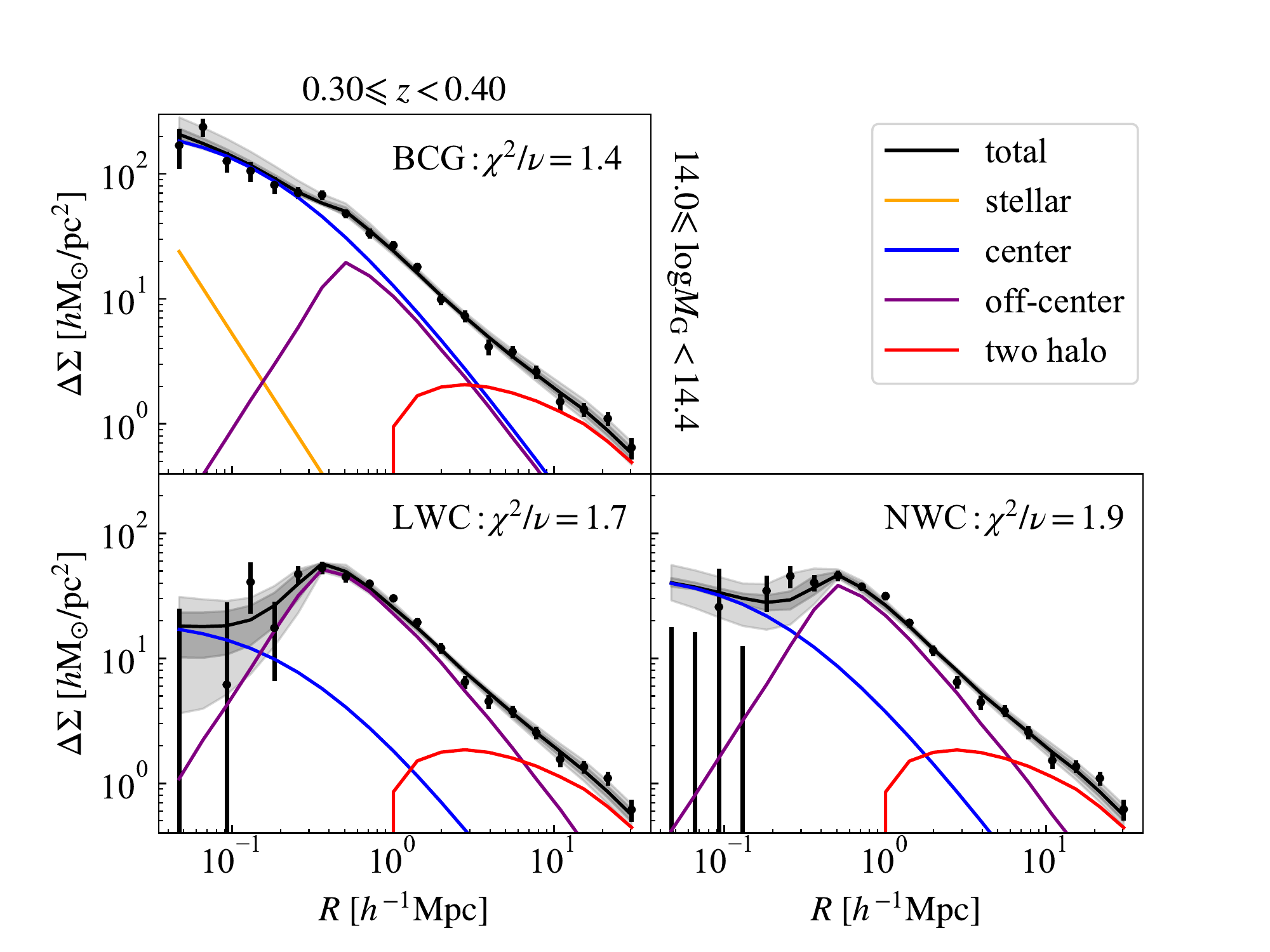}
    \caption{The best-fit results and different model components for lens sample with mass $14.0 \leqslant \log M_{\rm G} < 14.4$ at redshift $0.3 \leqslant z < 0.4$. The dark and light grey shaded regions represent the 68 and 95 confidence intervals of the total ESD fitting results, respectively. The three panels correspond to the three different type of group centers. The different colored lines represent the halo components labelled in the legend. The $\chi^2 / \nu$ value is also provided in each panel. Note here the degrees of freedom has a value of $\nu=15$. }
    \label{fig:fitted}
\end{figure*}

\subsection{Model fitting}

For each of our 35 lens samples, we use a Gaussian likelihood function with covariance matrix built from 400 jackknife samplings,
\begin{equation}
\ln\mathcal{L}(\mathbf{X}|\mathbf{\Theta})=
-0.5((\mathbf{X}-\mathbf{ESD_{m}})^TC^{-1}(\mathbf{X}-\mathbf{ESD_{m}}))\,,
\end{equation}
where $\mathbf{X}$ is the ESD observational measurement data vector, $\mathbf{ESD_{m}}$ is the
model and $C^{-1}$ is the inverse of the covariance
matrix. $\mathbf{\Theta}$ denotes the parameters in the ESD model.
In order to minimize the prior influence, we use broad flat priors for
all of the model parameters. We set the logarithm of halo mass $\log M_{\rm h}$ to be within $(12.0, 16.0)$, concentration range $(1.0, 20.0)$, central fraction $f_{\rm cen}$ to be within $(0.0, 1.0)$,  $R_{\rm sig}/r_{\rm 180m}$
range $(0.0, 1.0)$ and bias $b_{\rm h}$ range $(0.3, 10.0)$.
The halo parameters are estimated by the Markov Chain Monte Carlo fitting method using the emcee package\footnote{\url{https://emcee.readthedocs.io/en/stable}} \citep{Foreman_Mackey_2013}. For each of the lens samples, we adopt an ensemble sampler with 72 walkers over a chain of 720000 steps, where the first 72000 steps are discarded in our subsequent analysis.
%Once all these chains are generated, we combine them together and thin them by a factor of 10. 
From each of the final chain, we take the set of parameters with the least $\chi^2$ value as the best-fit results. The confidence regions of these parameters are obtained from their marginalized distributions.

For those who are interested in the whole list of parameters, we provide in the upper part of Table \ref{tab:pameters} with the best-fit parameters for all the 35 lens samples in different mass and redshift bins. For simplicity, here we only provide the best-fit parameters and $\chi^2$ values for the BCG group centers. 
%While the {\it averaged} values over the three centroid schemes are provided for the halo mass, concentration and bias of the all the lens samples, $\log <M_h>$, $<c>$, $<b_h>$. 
%Their error bars are calculated by taking the maximum of their fitting errors and the scatter among them.
As an illustration, in each panel of Fig. \ref{fig:esds} we show the best-fit model predictions of the ESD for the corresponding BCGs centroid lens sample as solid lines. 

%Similar to Fig. \ref{fig:fitted}, the model predictions show nice agreement with the ESD observational over a large scale range. \mf{I feel that this paragraph should go somewhere else, like maybe after the ECDs plotted as BCG, LWC, and NWC (showing the four components)}

To be more specific, we show in Fig. \ref{fig:fitted} the best-fit results for a typical lens sample as shown in the center of Fig. \ref{fig:esds}, 
i.e., groups with mass $14.0 \leqslant \log M_{\rm G} < 14.4$ at redshift $0.3 \leqslant z < 0.4$. 
Here results are shown for three different types of group center as indicated in different panels. 
In each panel, the total and different ESD components as described in Eq. \ref{eq:esdmodel} are presented using different style lines. 
%\mf{I edited the following sentences a lot, but I'm pretty sure this is what you mean:}
For all cases the four component model fits well on all scales.
%The four-components model fits well on the whole scale for all the centroids. 
The stellar component contributes mainly at small scales for the BCG centroid (note that we do not include the stellar component for LWC and NWC centroids);
the central term dominates on small scales;
the off-center term has relatively more impact on intermediate scales;%dominates over the central term on intedmediate scales for LWC and NWC centroids, yet has roughly equal contribution for BCG centroids.
and the two halo term dominates on large scales.%, which provides good halo bias estimations. 

\begin{figure*}
    \centering
    \includegraphics[width=0.6\textwidth]{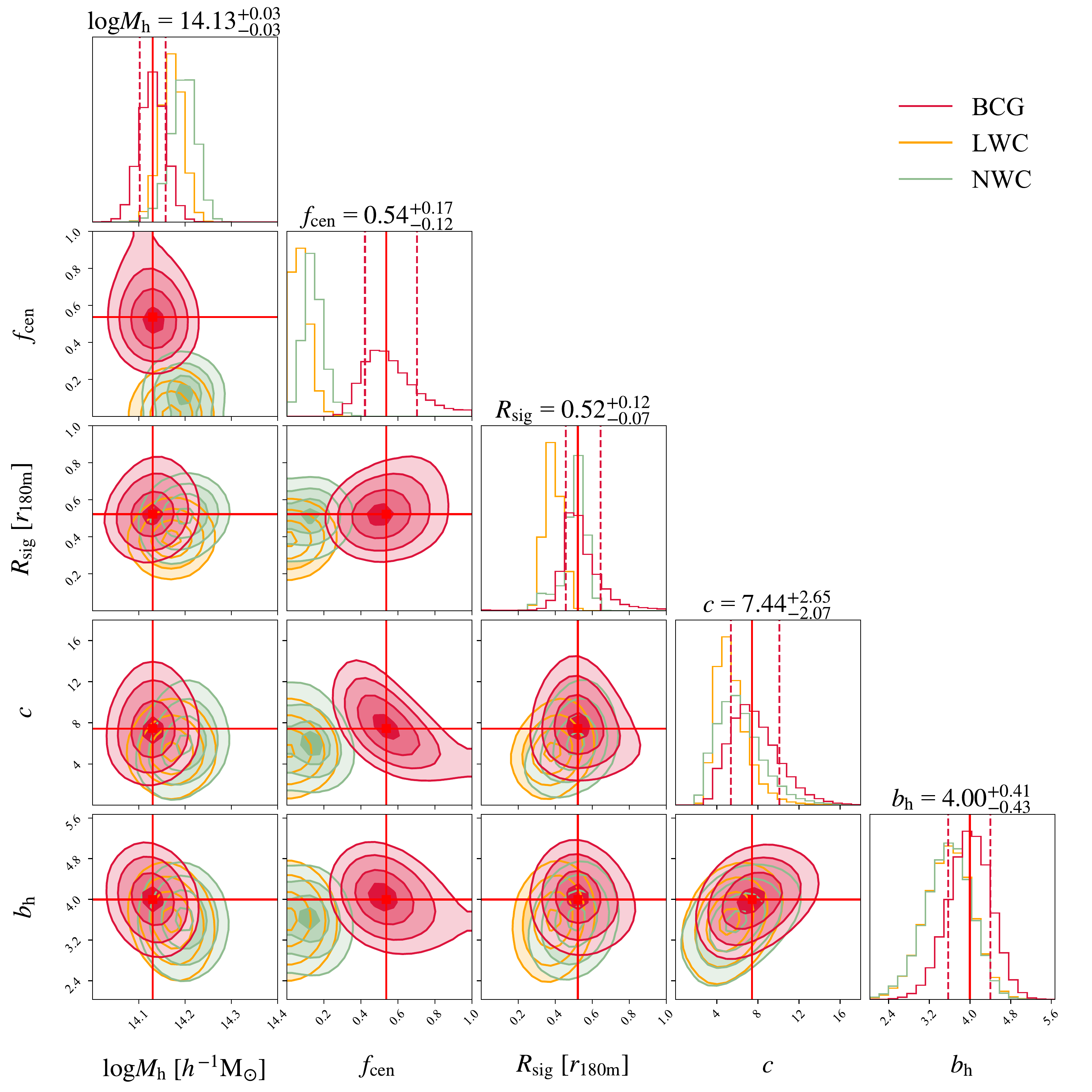}
    \caption{An example of marginalized posterior
distributions of the five parameters $\log M_{\rm h}$, $R_{\rm sig}$, $f_{\rm cen}$, $c$ and $b_{\rm h}$ for the lens sample with mass $14.0 \leqslant \log M_{\rm G} < 14.4$ at redshift $0.3\leqslant z < 0.4$. Here the red, orange and green contours show the result for BCG, LWC, NWC centroids, respectively.
The red solid lines show the best-fit result for the BCG sample. The red dashed lines show the 1-$\sigma$ confidence interval. 
}
    \label{fig:posterior}
\end{figure*}

Fig. \ref{fig:posterior} shows the marginalized posterior
distributions of the five parameters for ESDs in the above sample for the three group centers (see the bottom part of Table \ref{tab:pameters} for the detailed values of the best-fit parameters). The general properties of these five parameters are summarized below: 

\begin{itemize}
\item The average halo mass, $M_{\rm h}$, for each group centroid can be precisely measured at about 3-4\% level, while the difference among different centroids is less than 5\% level. Taking the statistical error and systematic difference into account, our halo mass estimation for this lens sample can achieve a precision at about 5\% level.

\item The concentrations $c$ of the dark matter halos can also be well constrained and agree with each other among three centroids. Taking into account the significantly different ESD behaviors at small scales, the nice agreement of $c$ is not trivial and can be regarded as a demonstration of the self-consistency of our model constraints. For the BCG case, we also see that the concentration $c$ is somewhat degenerate with $f_{\rm cen}$, where a smaller $c$ is associated with a larger $f_{\rm cen}$. 

\item The $b_{\rm h}$ of the three centroids agrees with each other well within the 1-$\sigma$ error bar. This agreement is not surprising, as the ESDs at large scales are almost the same for these three group centroids.

\item Within those five free parameters, we see the center fraction $f_{\rm cen}$ has the biggest difference among the three different group centroids. The BCG centroid has a $f_{\rm cen}\sim 0.55$ and the LWC and NWC centroid have much lower $f_{\rm cen}$ values. Thus, overall, BCG is the best tracer of the halo center. Interestingly, such behavior is in good agreement with the related probes in literature \citep[e.g.][]{Skibba2011, Wang2014,Lange_2018}. For instance, by comparing the diffused X-ray peak positions and the most massive galaxies (MMGs) in the groups, \citet{Wang2014} found that $\sim 65\%$ MMGs are located at the X-ray peak positions.

\item Although the center fraction $f_{\rm cen}$ for different group centroids are quite different, the off-center parameters $R_{\rm sig}$ are not that different among these schemes. And it is also quite evident that Rayleigh distribution can nicely describe the off-centering effect, in its ability to reproduce the subtle ESD features at small scales, as shown in Fig. \ref{fig:fitted}. 
\end{itemize}

According to the above inspections, we can see that the mass, concentration and bias of the halos obtained from the three different centroid schemes are in good agreement with each other. Furthermore, the 
difference seen in the $f_{\rm cen}$ parameters for the three different centroid schemes is also theoretically expected \citep[e.g.][]{Skibba2011, Wang2014, Lange_2018}. %\mf{cite}
Thus we believe that the main halo properties we obtain from our ESD measurements should be quite reliable.

\begin{table*}
\centering
 \caption{Posterior of parameters we obtained from the ESDs of our 35 lens samples (upper part) and the 20th lens sample with three types of group centers (lower part). In the upper part, we only list the posterior for the BCG group centers. In the bottom part, we list the results of all three centroids for sample 20, shown in Fig. \ref{fig:fitted}. The redshift/mass range refers to the redshift/log of prior group mass used in the bin. The following five columns $\log M_{\rm h}$, $c$, $b_{\rm h}$, $f_{\rm cen}$ and $R_{\rm sig}$ are the posterior of the parameters obtained from the MCMC chains. 
 The upper and lower 1-$\sigma$ errors are given using the superscript and subscript, respectively. We also give the $\chi^2 / \nu$ showing the fitting quality for each sample (here the degree of freedom of the system is either 14 or 15, depending on the redshift bin used (see Section 3.1)). %\mf{Not necessary but this caption could be cleaned up a bit. Like maybe talk about upper and lower part first, then you'll only need to describe each column afterwards.}
}
 \label{tab:pameters}
\begin{tabular}{l c c c c c c c c c}
 \hline\hline
  Sample & mass range & redshift range & $\log M_{\rm h}$ & $f_{\rm cen}$ & $R_{\rm sig}$ &  $c$ & $b_{\rm h}$ & $ \chi^2/\nu$ & \\ \hline
1 &
[13.00,13.20) &
[0.10,0.20) & 
$ 13.28 ^{+ 0.03 }_{- 0.03 } $ &
$ 0.53 ^{+ 0.10 }_{- 0.07 } $ &
$ 0.50 ^{+ 0.07 }_{- 0.06 } $ &
$ 14.86 ^{+ 2.95 }_{- 3.81 } $ &
$ 1.84 ^{+ 0.20 }_{- 0.20 } $ &
1.10
\\
2 &
[13.20,13.40) &
[0.10,0.20) & 
$ 13.44 ^{+ 0.05 }_{- 0.05 } $ &
$ 0.67 ^{+ 0.20 }_{- 0.17 } $ &
$ 0.68 ^{+ 0.33 }_{- 0.25 } $ &
$ 7.25 ^{+ 3.37 }_{- 2.08 } $ &
$ 1.48 ^{+ 0.24 }_{- 0.24 } $ &
1.05
\\
3 &
[13.40,13.60) &
[0.10,0.20) & 
$ 13.70 ^{+ 0.04 }_{- 0.04 } $ &
$ 0.43 ^{+ 0.12 }_{- 0.08 } $ &
$ 0.53 ^{+ 0.08 }_{- 0.06 } $ &
$ 11.78 ^{+ 3.86 }_{- 3.71 } $ &
$ 1.30 ^{+ 0.31 }_{- 0.32 } $ &
1.41
\\
4 &
[13.60,13.80) &
[0.10,0.20) & 
$ 13.86 ^{+ 0.05 }_{- 0.05 } $ &
$ 0.62 ^{+ 0.23 }_{- 0.17 } $ &
$ 0.64 ^{+ 0.30 }_{- 0.18 } $ &
$ 6.42 ^{+ 3.66 }_{- 2.13 } $ &
$ 1.95 ^{+ 0.43 }_{- 0.44 } $ &
1.80
\\
5 &
[13.80,14.00) &
[0.10,0.20) & 
$ 14.12 ^{+ 0.04 }_{- 0.04 } $ &
$ 0.32 ^{+ 0.05 }_{- 0.04 } $ &
$ 0.67 ^{+ 0.05 }_{- 0.04 } $ &
$ 13.31 ^{+ 1.50 }_{- 1.94 } $ &
$ 2.36 ^{+ 0.51 }_{- 0.50 } $ &
1.45
\\
6 &
[14.00,14.40) &
[0.10,0.20) & 
$ 14.27 ^{+ 0.06 }_{- 0.05 } $ &
$ 0.72 ^{+ 0.19 }_{- 0.17 } $ &
$ 0.88 ^{+ 0.48 }_{- 0.45 } $ &
$ 4.01 ^{+ 1.68 }_{- 1.02 } $ &
$ 2.22 ^{+ 0.69 }_{- 0.68 } $ &
1.71
\\
7 &
[14.40,15.00) &
[0.10,0.20) & 
$ 14.71 ^{+ 0.04 }_{- 0.05 } $ &
$ 0.35 ^{+ 0.07 }_{- 0.05 } $ &
$ 0.90 ^{+ 0.12 }_{- 0.08 } $ &
$ 9.09 ^{+ 1.65 }_{- 1.99 } $ &
$ 3.73 ^{+ 1.08 }_{- 1.09 } $ &
2.01
\\
8 &
[13.00,13.20) &
[0.20,0.30) & 
$ 13.15 ^{+ 0.04 }_{- 0.04 } $ &
$ 0.50 ^{+ 0.11 }_{- 0.07 } $ &
$ 0.54 ^{+ 0.12 }_{- 0.08 } $ &
$ 13.33 ^{+ 3.59 }_{- 3.55 } $ &
$ 1.73 ^{+ 0.12 }_{- 0.12 } $ &
1.97
\\
9 &
[13.20,13.40) &
[0.20,0.30) & 
$ 13.37 ^{+ 0.03 }_{- 0.03 } $ &
$ 0.40 ^{+ 0.06 }_{- 0.04 } $ &
$ 0.52 ^{+ 0.05 }_{- 0.04 } $ &
$ 16.27 ^{+ 2.01 }_{- 3.18 } $ &
$ 1.56 ^{+ 0.14 }_{- 0.14 } $ &
2.73
\\
10 &
[13.40,13.60) &
[0.20,0.30) & 
$ 13.49 ^{+ 0.04 }_{- 0.04 } $ &
$ 0.62 ^{+ 0.16 }_{- 0.12 } $ &
$ 0.75 ^{+ 0.22 }_{- 0.21 } $ &
$ 7.88 ^{+ 2.79 }_{- 2.18 } $ &
$ 1.78 ^{+ 0.18 }_{- 0.19 } $ &
1.80
\\
11 &
[13.60,13.80) &
[0.20,0.30) & 
$ 13.72 ^{+ 0.04 }_{- 0.04 } $ &
$ 0.64 ^{+ 0.18 }_{- 0.14 } $ &
$ 0.59 ^{+ 0.19 }_{- 0.23 } $ &
$ 7.59 ^{+ 2.63 }_{- 2.01 } $ &
$ 2.25 ^{+ 0.25 }_{- 0.25 } $ &
2.52
\\
12 &
[13.80,14.00) &
[0.20,0.30) & 
$ 13.90 ^{+ 0.03 }_{- 0.03 } $ &
$ 0.50 ^{+ 0.14 }_{- 0.10 } $ &
$ 0.56 ^{+ 0.15 }_{- 0.07 } $ &
$ 9.44 ^{+ 3.30 }_{- 2.53 } $ &
$ 2.34 ^{+ 0.30 }_{- 0.31 } $ &
1.62
\\
13 &
[14.00,14.40) &
[0.20,0.30) & 
$ 14.20 ^{+ 0.03 }_{- 0.03 } $ &
$ 0.48 ^{+ 0.08 }_{- 0.07 } $ &
$ 0.78 ^{+ 0.06 }_{- 0.06 } $ &
$ 9.11 ^{+ 1.91 }_{- 1.76 } $ &
$ 3.65 ^{+ 0.34 }_{- 0.35 } $ &
1.89
\\
14 &
[14.40,15.00) &
[0.20,0.30) & 
$ 14.58 ^{+ 0.04 }_{- 0.04 } $ &
$ 0.74 ^{+ 0.18 }_{- 0.18 } $ &
$ 0.69 ^{+ 0.60 }_{- 0.27 } $ &
$ 3.44 ^{+ 1.36 }_{- 0.78 } $ &
$ 5.01 ^{+ 0.96 }_{- 1.04 } $ &
1.47
\\
15 &
[13.00,13.20) &
[0.30,0.40) & 
$ 13.15 ^{+ 0.04 }_{- 0.04 } $ &
$ 0.60 ^{+ 0.11 }_{- 0.08 } $ &
$ 0.75 ^{+ 0.14 }_{- 0.13 } $ &
$ 10.90 ^{+ 2.18 }_{- 2.26 } $ &
$ 1.67 ^{+ 0.11 }_{- 0.11 } $ &
1.46
\\
16 &
[13.20,13.40) &
[0.30,0.40) & 
$ 13.37 ^{+ 0.04 }_{- 0.04 } $ &
$ 0.52 ^{+ 0.14 }_{- 0.09 } $ &
$ 0.65 ^{+ 0.09 }_{- 0.08 } $ &
$ 9.73 ^{+ 2.71 }_{- 2.73 } $ &
$ 1.77 ^{+ 0.14 }_{- 0.14 } $ &
1.21
\\
17 &
[13.40,13.60) &
[0.30,0.40) & 
$ 13.51 ^{+ 0.03 }_{- 0.04 } $ &
$ 0.48 ^{+ 0.12 }_{- 0.08 } $ &
$ 0.66 ^{+ 0.11 }_{- 0.09 } $ &
$ 10.05 ^{+ 2.79 }_{- 2.75 } $ &
$ 2.01 ^{+ 0.18 }_{- 0.19 } $ &
1.57
\\
18 &
[13.60,13.80) &
[0.30,0.40) & 
$ 13.70 ^{+ 0.04 }_{- 0.04 } $ &
$ 0.66 ^{+ 0.20 }_{- 0.17 } $ &
$ 0.67 ^{+ 0.41 }_{- 0.22 } $ &
$ 5.41 ^{+ 2.46 }_{- 1.44 } $ &
$ 2.38 ^{+ 0.25 }_{- 0.24 } $ &
2.29
\\
19 &
[13.80,14.00) &
[0.30,0.40) & 
$ 13.84 ^{+ 0.05 }_{- 0.05 } $ &
$ 0.59 ^{+ 0.18 }_{- 0.13 } $ &
$ 0.79 ^{+ 0.27 }_{- 0.21 } $ &
$ 7.26 ^{+ 2.71 }_{- 2.17 } $ &
$ 2.93 ^{+ 0.31 }_{- 0.31 } $ &
1.86
\\
20 &
[14.00,14.40) &
[0.30,0.40) & 
$ 14.13 ^{+ 0.03 }_{- 0.03 } $ &
$ 0.54 ^{+ 0.17 }_{- 0.12 } $ &
$ 0.52 ^{+ 0.12 }_{- 0.07 } $ &
$ 7.44 ^{+ 2.65 }_{- 2.07 } $ &
$ 3.81 ^{+ 0.39 }_{- 0.41 } $ &
1.38
\\
21 &
[14.40,15.00) &
[0.30,0.40) & 
$ 14.59 ^{+ 0.04 }_{- 0.04 } $ &
$ 0.47 ^{+ 0.16 }_{- 0.10 } $ &
$ 0.62 ^{+ 0.12 }_{- 0.11 } $ &
$ 6.01 ^{+ 2.49 }_{- 1.89 } $ &
$ 6.07 ^{+ 1.02 }_{- 1.10 } $ &
1.45
\\
22 &
[13.00,13.20) &
[0.40,0.50) & 
$ 13.07 ^{+ 0.04 }_{- 0.04 } $ &
$ 0.58 ^{+ 0.15 }_{- 0.10 } $ &
$ 0.55 ^{+ 0.14 }_{- 0.08 } $ &
$ 12.82 ^{+ 3.89 }_{- 4.59 } $ &
$ 1.74 ^{+ 0.12 }_{- 0.12 } $ &
2.45
\\
23 &
[13.20,13.40) &
[0.40,0.50) & 
$ 13.26 ^{+ 0.05 }_{- 0.05 } $ &
$ 0.67 ^{+ 0.15 }_{- 0.11 } $ &
$ 0.80 ^{+ 0.18 }_{- 0.19 } $ &
$ 9.29 ^{+ 2.30 }_{- 2.46 } $ &
$ 2.00 ^{+ 0.16 }_{- 0.16 } $ &
2.66
\\
24 &
[13.40,13.60) &
[0.40,0.50) & 
$ 13.44 ^{+ 0.05 }_{- 0.05 } $ &
$ 0.57 ^{+ 0.22 }_{- 0.13 } $ &
$ 0.73 ^{+ 0.28 }_{- 0.19 } $ &
$ 7.31 ^{+ 3.56 }_{- 2.82 } $ &
$ 1.95 ^{+ 0.19 }_{- 0.20 } $ &
1.48
\\
25 &
[13.60,13.80) &
[0.40,0.50) & 
$ 13.61 ^{+ 0.04 }_{- 0.05 } $ &
$ 0.48 ^{+ 0.14 }_{- 0.09 } $ &
$ 0.63 ^{+ 0.13 }_{- 0.10 } $ &
$ 10.34 ^{+ 3.08 }_{- 3.21 } $ &
$ 2.56 ^{+ 0.25 }_{- 0.25 } $ &
2.24
\\
26 &
[13.80,14.00) &
[0.40,0.50) & 
$ 13.77 ^{+ 0.05 }_{- 0.04 } $ &
$ 0.63 ^{+ 0.21 }_{- 0.16 } $ &
$ 0.48 ^{+ 0.51 }_{- 0.10 } $ &
$ 9.55 ^{+ 5.76 }_{- 3.41 } $ &
$ 2.97 ^{+ 0.38 }_{- 0.39 } $ &
3.02
\\
27 &
[14.00,14.40) &
[0.40,0.50) & 
$ 14.13 ^{+ 0.03 }_{- 0.03 } $ &
$ 0.51 ^{+ 0.21 }_{- 0.12 } $ &
$ 0.63 ^{+ 0.13 }_{- 0.09 } $ &
$ 7.78 ^{+ 3.80 }_{- 2.93 } $ &
$ 4.20 ^{+ 0.49 }_{- 0.53 } $ &
1.29
\\
28 &
[14.40,15.00) &
[0.40,0.50) & 
$ 14.47 ^{+ 0.05 }_{- 0.04 } $ &
$ 0.64 ^{+ 0.24 }_{- 0.23 } $ &
$ 0.41 ^{+ 0.54 }_{- 0.17 } $ &
$ 5.72 ^{+ 2.80 }_{- 1.47 } $ &
$ 5.73 ^{+ 1.02 }_{- 1.05 } $ &
1.64
\\
29 &
[13.00,13.20) &
[0.50,0.60) & 
$ 12.93 ^{+ 0.06 }_{- 0.06 } $ &
$ 0.64 ^{+ 0.22 }_{- 0.23 } $ &
$ 0.49 ^{+ 0.23 }_{- 0.29 } $ &
$ 10.17 ^{+ 4.72 }_{- 3.47 } $ &
$ 1.60 ^{+ 0.16 }_{- 0.17 } $ &
1.45
\\
30 &
[13.20,13.40) &
[0.50,0.60) & 
$ 13.18 ^{+ 0.07 }_{- 0.06 } $ &
$ 0.60 ^{+ 0.21 }_{- 0.16 } $ &
$ 0.63 ^{+ 0.36 }_{- 0.21 } $ &
$ 8.82 ^{+ 4.40 }_{- 3.07 } $ &
$ 1.92 ^{+ 0.20 }_{- 0.20 } $ &
1.82
\\
31 &
[13.40,13.60) &
[0.50,0.60) & 
$ 13.41 ^{+ 0.06 }_{- 0.06 } $ &
$ 0.70 ^{+ 0.18 }_{- 0.17 } $ &
$ 0.72 ^{+ 0.44 }_{- 0.36 } $ &
$ 7.38 ^{+ 3.37 }_{- 2.14 } $ &
$ 2.19 ^{+ 0.24 }_{- 0.25 } $ &
1.26
\\
32 &
[13.60,13.80) &
[0.50,0.60) & 
$ 13.59 ^{+ 0.06 }_{- 0.06 } $ &
$ 0.27 ^{+ 0.10 }_{- 0.08 } $ &
$ 0.54 ^{+ 0.09 }_{- 0.08 } $ &
$ 12.97 ^{+ 3.64 }_{- 3.93 } $ &
$ 2.61 ^{+ 0.31 }_{- 0.31 } $ &
1.85
\\
33 &
[13.80,14.00) &
[0.50,0.60) & 
$ 13.65 ^{+ 0.06 }_{- 0.06 } $ &
$ 0.25 ^{+ 0.36 }_{- 0.16 } $ &
$ 0.28 ^{+ 0.10 }_{- 0.07 } $ &
$ 9.25 ^{+ 6.01 }_{- 4.19 } $ &
$ 3.40 ^{+ 0.47 }_{- 0.50 } $ &
1.05
\\
34 &
[14.00,14.40) &
[0.50,0.60) & 
$ 14.15 ^{+ 0.04 }_{- 0.04 } $ &
$ 0.41 ^{+ 0.12 }_{- 0.07 } $ &
$ 0.65 ^{+ 0.09 }_{- 0.06 } $ &
$ 11.53 ^{+ 2.87 }_{- 3.70 } $ &
$ 3.85 ^{+ 0.60 }_{- 0.63 } $ &
2.31
\\
35 &
[14.40,15.00) &
[0.50,0.60) & 
$ 14.59 ^{+ 0.09 }_{- 0.09 } $ &
$ 0.60 ^{+ 0.24 }_{- 0.20 } $ &
$ 1.09 ^{+ 1.03 }_{- 0.35 } $ &
$ 4.07 ^{+ 3.14 }_{- 1.56 } $ &
$ 3.65 ^{+ 1.66 }_{- 1.46 } $ &
0.65
\\
\hline \\
\hline\hline

20 - BCG &
[14.00,14.40) &
[0.30,0.40) & 
$ 14.13 ^{+ 0.03 }_{- 0.03 } $ &
$ 0.54 ^{+ 0.17 }_{- 0.12 } $ &
$ 0.52 ^{+ 0.12 }_{- 0.07 } $ &
$ 7.44 ^{+ 2.65 }_{- 2.07 } $ &
$ 3.81 ^{+ 0.39 }_{- 0.41 } $ &
1.38
\\
20 - LWC &
[14.00,14.40) &
[0.30,0.40) & 
$ 14.17 ^{+ 0.02 }_{- 0.02 } $ &
$ 0.07 ^{+ 0.05 }_{- 0.04 } $ &
$ 0.39 ^{+ 0.05 }_{- 0.04 } $ &
$ 5.21 ^{+ 1.75 }_{- 1.17 } $ &
$ 3.42 ^{+ 0.42 }_{- 0.45 } $ &
1.66
\\
20 - NWC &
[14.00,14.40) &
[0.30,0.40) & 
$ 14.20 ^{+ 0.03 }_{- 0.03 } $ &
$ 0.14 ^{+ 0.06 }_{- 0.05 } $ &
$ 0.51 ^{+ 0.05 }_{- 0.06 } $ &
$ 5.97 ^{+ 2.44 }_{- 1.90 } $ &
$ 3.42 ^{+ 0.41 }_{- 0.45 } $ &
1.89
\\
\hline
\end{tabular}
\end{table*}

\section{Halo properties}

\label{sec_properties}
Having demonstrated that the model parameters can be reliably constrained using the ESD measurements, we proceed to discuss the halo properties we extracted. As different group centers may impact the $f_{\rm cen}$ and $R_{\rm sig}$ parameters, we focus only on the other three parameters obtained from the BCG centroid schemes. For the readers who are interested in the actual differences among the three group centroid schemes, we include a discussion in the Appendix \ref{AP:THREE} (Fig. \ref{fig:halo_prop2}).
%could refer to the results shown in Fig. of the Appendix.

\begin{figure*}[!]
    \centering
    \includegraphics[width=1\textwidth]{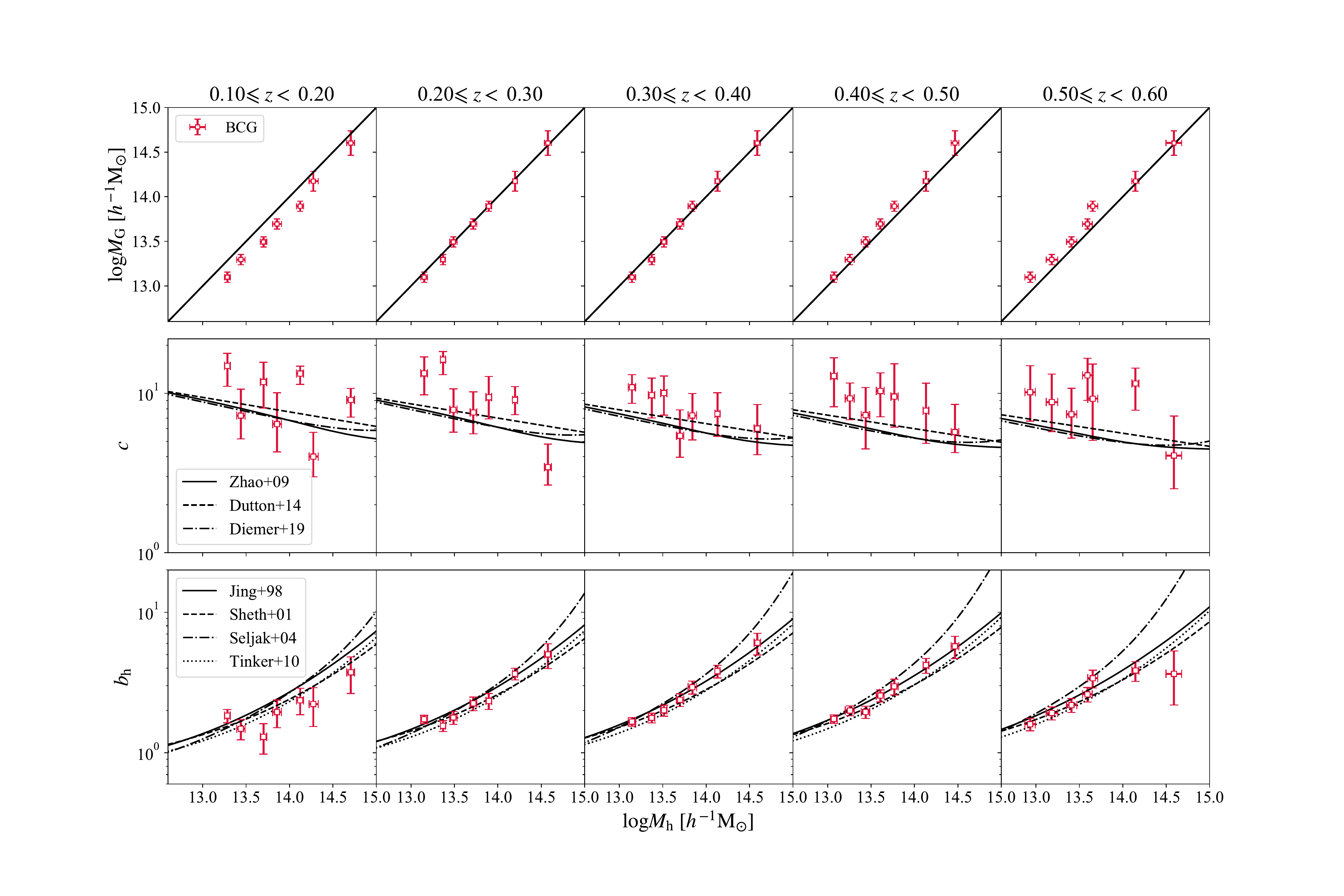}
    \caption{ 
    This plot shows the halo properties measured from the ESDs of our 35 lens samples. Shown in different columns are results for lens samples in different redshift bins as indicated on top of each column. The $x$-axis is the halo mass measured from weak lensing. The data points in each panel show the best-fit value for the BCG centroid schemes. %average values obtained from three centroid schemes. 
    (a) Shown in the top row panels are the comparisons between the lensing halo mass ($\log M_{\rm h}$) v.s. the group mass estimated using the abundance matching method ($\log M_{\rm G}$). Note that the error along the $y$-axis is the standard error of the group mass of the sample, while the error along the $x$-axis indicate the 1-$\sigma$ confidence interval of the fitting. The black lines are the reference lines where $\log M_{\rm G} = \log M_{\rm h}$.
    (b) The middle row panels show the concentration-mass relation measured from the ESDs. The $y$-axis is the halo concentration. The theoretical predictions of the c-M relations from \citet{Zhao2009, Dutton2014, Diemer2019} are shown as the solid, dashed and dashdot lines, respectively. 
    (c) The bottom row panels show the halo bias-mass relations. 
    The theoretical predictions of the b-M relations %\mf{is this supposed to be c-M?} 
    from \citet{Jing1998,Sheth2001,Seljak2004,Tinker2010} are shown as the solid, dashed, dashdot and dotted lines, respectively.  %Except for the lowest redshift bin, our measurements in general have very good agreement with those theoretical model  predictions.
    }
    \label{fig:halo_prop}
\end{figure*}

\subsection{Comparing the halo mass and group mass}

As we can see from the likelihood distribution of parameters in Fig. \ref{fig:posterior}, we can have fairly good constraints on the halo bias, concentration and mass of our lens samples.
It would be interesting to have a direct comparison between the halo masses obtained from the ESD fittings and the group masses provided by the group catalog. As the group mass obtained in Y21 is based on the halo abundance matching method, it in general refers to the abundance of the groups. Provided that the cosmologies are correct, the group mass should not have systematical biases. %\mf{I think you are using the same cosmologies, and the masses should match if the cosmologies match? I haven't tested this for all cosmologies, but I think the HMF in different cosmologies match in mock WL simulations/papers}

In the upper row of Fig. \ref{fig:halo_prop}, we compare the best-fit halo masses obtained from the weak lensing signals and the mean group masses obtained from the abundance matching method. The two agree with each other very well in the redshift range $0.2 \leqslant z <0.4$, but are systematically different at lower and higher redshift bins. In the lowest redshift bin, the group masses are systematically smaller than the halo masses, while in the two highest redshift bins, the group masses exceed the halo masses. According to Table \ref{tab:pameters}, the halo masses obtained from the ESDs are higher/lower than those obtained from the group catalog by about 0.1$\sim$0.2 dex in the lowest/highest redshift bins. This discrepancy is discussed in more detail with the help of cumulative halo mass functions in Section \ref{sec:CHMF_Y21}.

\begin{figure*}
    \centering
    \includegraphics[width=0.80\textwidth]{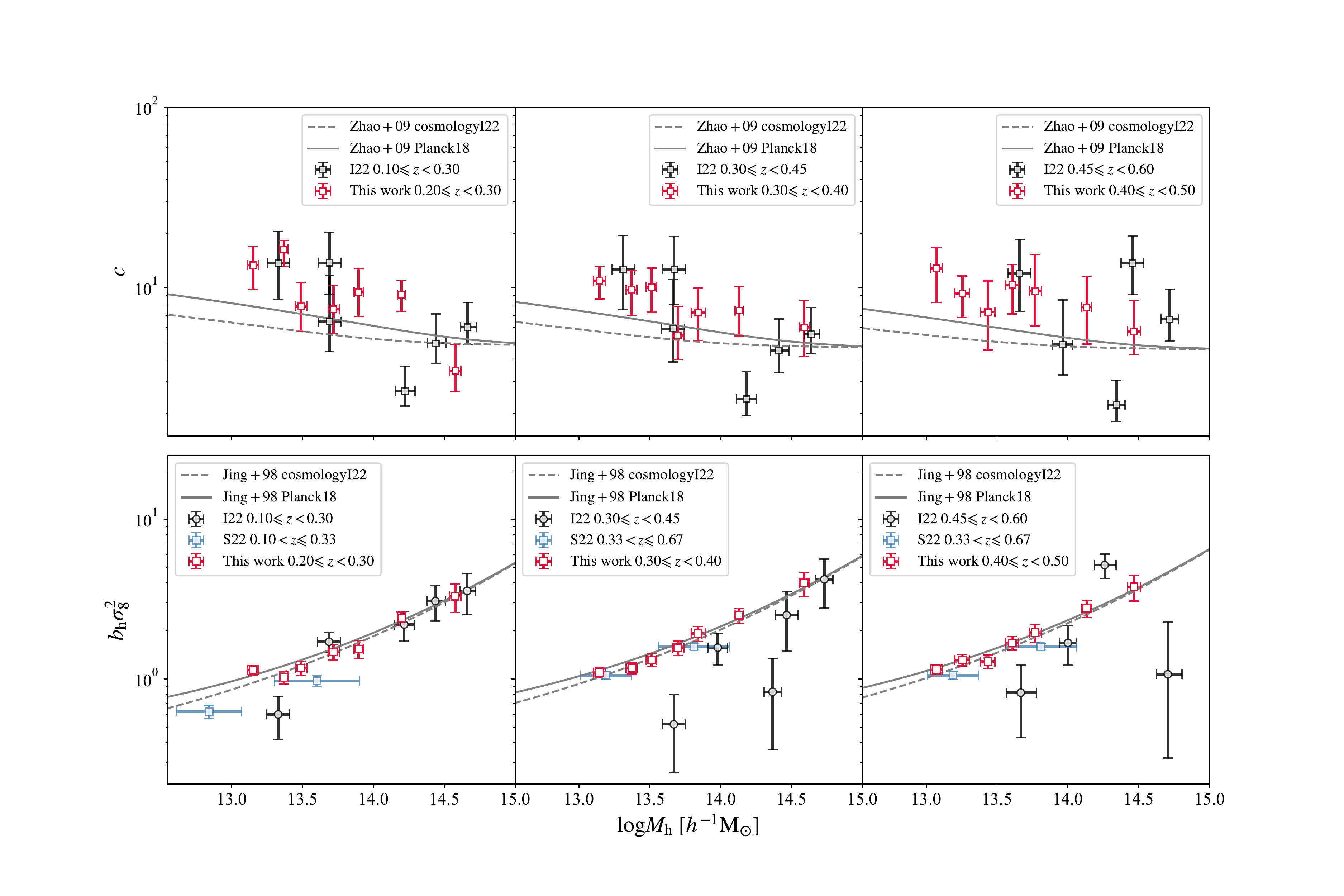}
    \caption{We compare our concentration and bias measurements to previous observations by \citet{Ingoglia_2022} and \citet{Sun2022} using different types of symbols, with average redshift of each sample within $0.2 \leqslant z <0.6$. Here we only show the results in three redshift bin within $0.2 \leqslant z <0.5$. We present the theoretical predictions under Planck18 cosmology in solid lines and the one under the cosmology obtained in I22 in dashed lines.
    }
    \label{fig:bias_ob}
\end{figure*}

\subsection{Concentration - halo mass relation}

In the middle row of Fig. \ref{fig:halo_prop}, we show the concentration of our lens samples as a function of halo mass. Although the error bars are somewhat large, there is a clear trend that the concentration decreases with the halo mass. 

For comparison, we also show in each panel the model predictions of \cite{Zhao2009,Dutton2014,Diemer2019} under the Planck18 cosmology in the related redshift bin. The corresponding results are shown as the solid, dashed and dot-dashed lines, respectively. Here we have properly converted their halo mass definition into ours, i.e., halos are defined to have 180 times the mean background density, using  the \textsc{colossus} package~\citep{Diemer2018}.
Overall, our observational constraints are somewhat higher than the model predictions, especially in the high redshift bins. %But take caution that there is some degeneracy between our model constraints on the concentration and fraction of centrals.  
Due to the large error bars, the mild redshift dependence presented in these theoretical models can not be verified.  

In Fig.\ref{fig:bias_ob} we also provide a comparison of concentration measurements with the data recently obtained by \citet[][hereafter I22]{Ingoglia_2022} from the ESDs around the AMICO clusters. 
% cosmology: 
Note that the cosmology adopted in that work is somewhat different from ours, i.e. a flat $\Lambda$CDM model with $\Omega_{\rm m} = 0.3$, $\Omega_{\rm b} = 0.05$, $h=0.7$ and the $\sigma_8$ they constrained is $0.63^{+0.11}_{-0.10}$.
The results are shown in the upper panels of Fig. \ref{fig:bias_ob}. Since I22 has less redshift bins, we only show our results in the middle three redshift bins and compare them with the closest redshift bins. In each panel, we show the theoretical predictions from \citet{Zhao2009} for Planck18 (solid curves) as well as the cosmology obtained by I22 (dashed lines). The two different sets of cosmology lead to slight differences in the theoretical predictions. Overall, our results seems to have a better quality than that of I22. But there is no indication for a systematic difference between our concentration measurement and those of I22.
%Given the quite large error bars, although somewhat higher than the theoretical predictions, the concentration obtained here  need more careful treatments before it can be used to further constrain the properties of the dark matter.

\subsection{Bias - halo mass relation}

In the lowest panels of Fig. \ref{fig:halo_prop}, we show the halo biases of our 35 lens samples as functions of the halo mass. 
For comparisons, in each redshift bin, we also show the theoretical model predictions of \citet[][]{Jing1998,Sheth2001,Seljak2004,Tinker2010}, presented using solid, dashed, dot-dashed, and dotted lines respectively. 
Again, we have properly converted their halo mass definition into ours. Within these theoretical predictions, \citet[][]{Seljak2004} model has the strongest halo mass and redshift evolution dependence under the Planck18 cosmology. The other three models, although with slightly different amplitudes, have very similar dependence on the halo mass and redshift. 

Our observational measurements of the halo biases have a good agreement with these theoretical model predictions, except for the lowest redshift bin. Overall, our observational data favor the model predictions of \citet[][]{Jing1998,Sheth2001,Tinker2010}.
The biases in the lowest redshift bin are slightly lower than the theoretical predictions. However, due to the large error bars in this bin, the deviation is only at about 1-2 $\sigma$ level. 
% cosmology
Nevertheless, it is still worthwhile to check its significance and origin, either from the lensing signal measurements or the cosmological perspectives, which we will carry out in a separate paper.

We also provide a comparison of our bias measurements with two sets of very recent observational constraints in the bottom panels of Fig. \ref{fig:bias_ob}. The first set of bias measurements are obtained by \citet[][hereafter S22]{Sun2022} from the  cross-correlation between the DESI groups and the CMB lensing signals. The other set of bias measurements are obtained by I22.
The results are shown in different symbols. 
In each panel, we only show the theoretical predictions from \citet{Jing1998} at the average redshift of our sample under the Planck 18 cosmology (solid curves) as well as the cosmological model obtained by I22 (dashed curves), as reference.
The results obtained by S22 is in rough agreement with our measurements. On the other hand, the results of I22 are somewhat lower than our bias measurements in the two higher redshift bins. Their error bars are larger due to their much smaller lens sample sizes.

\begin{figure*}
    \centering
    \includegraphics[width=1\textwidth]{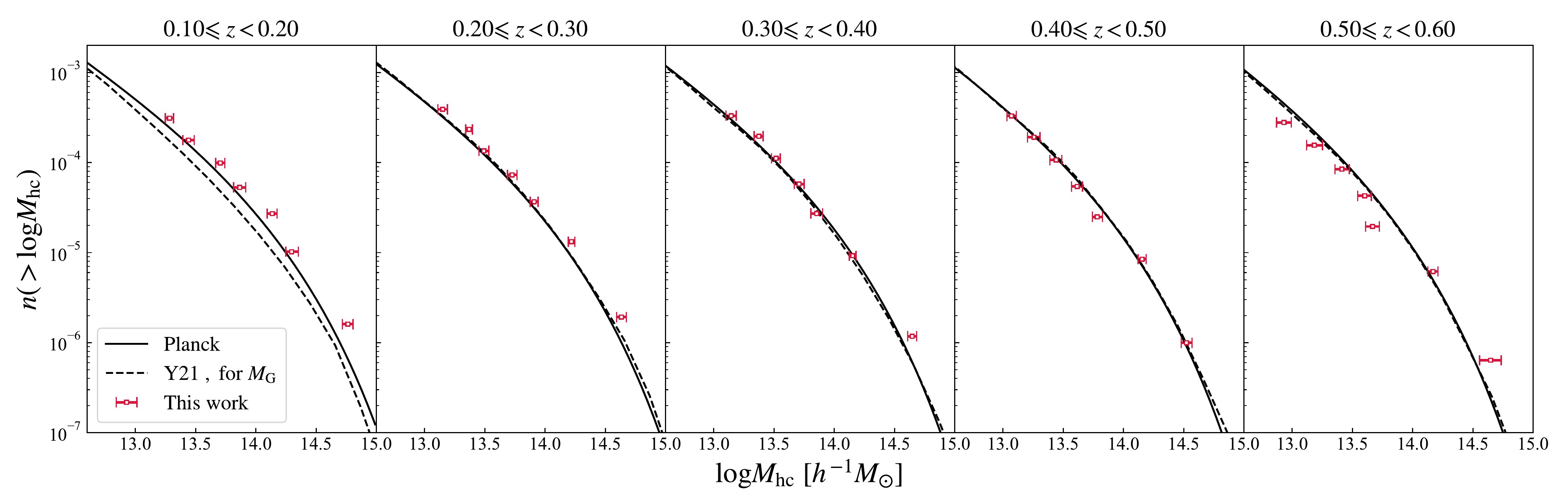}
    \caption{The cumulative halo mass functions (CHMFs) in five different redshift bins. The red points with error bars show the CHMFs obtained from the weak lensing ESD measurements. The black dashed lines show the CHMFs obtained from the group catalog of \cite{Yang2021} with the correction using Eq. \ref{eq:completeness}. The black solid lines show the theoretical CHMFs using fitting functions from \citet{Tinker2008} under the Planck18 cosmology \citep{2020A&A...641A...6P}. }
    \label{fig:chmf1}
\end{figure*}

\begin{table*}
\centering
 \caption{The values of cumulative halo mass function measured in this work (plotted as the red points in Fig. \ref{fig:chmf1}). There are seven data points in each redshift bin, provided within two columns: halo mass and the number density of halos with mass larger than this halo mass.}\label{tab:AHMF}
\begin{tabular}{cccccccccccccc}
\hline\hline
ID & \multicolumn{2}{c}{$0.1 \leqslant z< 0.2$} &  \multicolumn{2}{c}{$0.2 \leqslant z< 0.3$} & 
\multicolumn{2}{c}{$0.3 \leqslant z< 0.4$} & 
\multicolumn{2}{c}{$0.4 \leqslant z< 0.5$} & 
\multicolumn{2}{c}{$0.5 \leqslant z<  0.6$} &\\
 & $\log M_{\rm hc}$ & $n(> M_{\rm hc})$  & $\log M_{\rm hc}$ & $n(> M_{\rm hc})$ & $\log M_{\rm hc}$ & $n(> M_{\rm hc})$ & $\log M_{\rm hc}$ & $n(> M_{\rm hc})$ & $\log M_{\rm hc}$ & $n(> M_{\rm hc})$ \\
 & $\msunh$ &$(\times 10^{-5})$& $\msunh$ &$(\times 10^{-5})$& $\msunh$ &$(\times 10^{-5})$& $\msunh$ &$(\times 10^{-5})$& $\msunh$ &$(\times 10^{-5})$\\
 \hline
1
& $ 13.28 ^ { + 0.03 }_{- 0.03 } $ &  31.0
& $ 13.15 ^ { + 0.04 }_{- 0.04 } $ &  39.0
& $ 13.14 ^ { + 0.04 }_{- 0.04 } $ &  33.1
& $ 13.07 ^ { + 0.04 }_{- 0.04 } $ &  32.8
& $ 12.93 ^ { + 0.06 }_{- 0.06 } $ &  27.9
\\
2
& $ 13.44 ^ { + 0.05 }_{- 0.05 } $ &  17.7
& $ 13.37 ^ { + 0.03 }_{- 0.03 } $ &  23.3
& $ 13.37 ^ { + 0.04 }_{- 0.04 } $ &  19.5
& $ 13.26 ^ { + 0.05 }_{- 0.05 } $ &  19.1
& $ 13.18 ^ { + 0.07 }_{- 0.06 } $ &  15.6
\\
3
& $ 13.70 ^ { + 0.04 }_{- 0.04 } $ &  9.89
& $ 13.49 ^ { + 0.04 }_{- 0.04 } $ &  13.5
& $ 13.52 ^ { + 0.03 }_{- 0.04 } $ &  11.1
& $ 13.44 ^ { + 0.05 }_{- 0.05 } $ &  10.7
& $ 13.41 ^ { + 0.06 }_{- 0.06 } $ &  8.43
\\
4
& $ 13.86 ^ { + 0.05 }_{- 0.05 } $ &  5.29
& $ 13.73 ^ { + 0.04 }_{- 0.04 } $ &  7.29
& $ 13.71 ^ { + 0.04 }_{- 0.04 } $ &  5.76
& $ 13.62 ^ { + 0.04 }_{- 0.05 } $ &  5.41
& $ 13.60 ^ { + 0.06 }_{- 0.06 } $ &  4.27
\\
5
& $ 14.14 ^ { + 0.04 }_{- 0.04 } $ &  2.71
& $ 13.91 ^ { + 0.03 }_{- 0.03 } $ &  3.66
& $ 13.85 ^ { + 0.05 }_{- 0.05 } $ &  2.72
& $ 13.78 ^ { + 0.05 }_{- 0.04 } $ &  2.49
& $ 13.67 ^ { + 0.06 }_{- 0.06 } $ &  1.95
\\
6
& $ 14.30 ^ { + 0.06 }_{- 0.05 } $ &  1.02
& $ 14.22 ^ { + 0.03 }_{- 0.03 } $ &  1.32
& $ 14.15 ^ { + 0.03 }_{- 0.03 } $ &  0.922
& $ 14.16 ^ { + 0.03 }_{- 0.03 } $ &  0.845
& $ 14.17 ^ { + 0.04 }_{- 0.04 } $ &  0.617
\\
7
& $ 14.76 ^ { + 0.04 }_{- 0.05 } $ &  0.160
& $ 14.63 ^ { + 0.04 }_{- 0.04 } $ &  0.192
& $ 14.65 ^ { + 0.04 }_{- 0.04 } $ &  0.118
& $ 14.52 ^ { + 0.05 }_{- 0.04 } $ &  0.0994
& $ 14.65 ^ { + 0.09 }_{- 0.09 } $ &  0.0637
\\
\hline
\end{tabular}
\end{table*}

\section{Cumulative halo mass functions}
\label{sec:ahmf}
In Fig.\ref{fig:chmf1}, we show the cumulative halo mass functions (CHMFs) in five redshift bins obtained directly from the group catalog (the black dashed lines) and those from weak lensing measurement (the red points), and compare them to the theoretical predictions (the black solid lines).
%We first calculate CHMFs with completeness correction from the group catalog in  (the black dashed line in Fig. 7). 
In deriving the CHMFs, we have properly taken into account the completeness and purity corrections from the group catalog in Y21, as well as a halo mass correction due to the mass scatter in the weak lensing measurement. 
%Using the group abundance table generated in the previous step, we can calculate the corresponding effective group abundance for each stack bin used in this work.
%Combine the group abundance and the corrected halo mass, we can get the unbiased CHMFs (the red dots in Fig. 7).
In the rest of this section, we give some details regarding our data processing, and discuss about some of our limitations.
%and 

\subsection{CHMFs adopted in the group catalog}
\label{sec:CHMF_Y21}

For each of the redshift bins, we count the total number of groups that are more massive than a given group mass, $N_{\rm G}(> M_{\rm G})$. 
While in calculating this number, we have properly taken into account the completeness and purity of the groups as obtained by Y21 from the mock DESI redshift galaxy and group samples,
\begin{equation}
 N_{\rm G}(> M_{\rm G}) = \sum_i^{M_{{\rm G},i}> M_{\rm G}} f_{\rm group}(M_{{\rm G},i})/f_{\rm halo}(M_{{\rm G},i})\,,
 \label{eq:completeness}
\end{equation}
where $f_{\rm halo}$ and $f_{\rm group}$ are the completeness and purity of the groups (see Fig. 8 of Y21), respectively. %In practice, we take group mass $M_{{\rm G},i}$ instead of halo mass in the completeness $f_{\rm halo}$ to make the calculation.
Here $M_{{\rm G},i}$ is the mass of the i-th group in the whole catalog, after ranking according to their masses.
Note that since the group masses,
$M_{\rm G}$, provided in Y21 group catalog were assigned based on
abundance matching to the halo mass functions assuming the Planck18 cosmology, these masses are cosmology dependent. 
However, even if a different cosmology had been used to assign group masses, 
and $M_{\rm G}$ would be different, their rank-order would be preserved. Thus the ranked number of groups is cosmology independent.

Once we obtained the total number of groups $N_{\rm G}(> M_{\rm G})$ that are more massive than $M_{\rm G}$, we can normalize it according to the volume of each lens sample, and thus get the cumulative number density of the groups, $n_{\rm G}(> M_{\rm G})$, which are shown in Fig. \ref{fig:chmf1} as black dashed lines. 
Note that, in this step, since we need cosmological parameters, especially $\Omega_m$, to calculate the comoving volume, it is also
cosmology dependent. Here again, we adopt the parameters from the Planck18 cosmology for our distance calculation. 

Ideally, as the group masses, $M_{\rm G}$, were assigned according to the Planck18 cosmology, the dashed line in each panel of Fig. \ref{fig:chmf1} should agree with the theoretical prediction described by the solid line. However, we find that in the lowest redshift bin with $0.1\leqslant z <0.2$, it is somewhat lower than the expectation. This discrepancy indicates that the group abundance within this redshift bin is somewhat underestimated due to some systematic effects, such as redshift errors. Upon this feature, we are able to explain the discrepancy between the group mass and lensing mass shown in Fig. \ref{fig:halo_prop} with the following four main possible reasons: (1) the abundance of the groups is not consistent with Planck18 prediction in the lowest redshift bin, which was induced by different redshift bin widths that are used for group ranking (0.1 in this study v.s. 0.33 in Y21), (2) the Planck18 cosmology assumed in Y21 is incorrect, (3) there are 
systematic error in the halo masses inferred from the ESDs, and (4) the rank-order
according to total group luminosity is not equal to rank-order in
actual halo mass.

\subsection{CHMFs obtained from the weak lensing measurements}

\label{CHMF_lensing}
As we have obtained the measurements of the halo masses for each of our 35 lens samples, with a typical uncertainty at 4-13\% level (see Table \ref{tab:pameters}), it would be straightforward to measure the CHMFs. \citet[][hereafter D19]{Dong2019} pointed out that, the weak lensing mass measurement 
%inside group rank-order bins 
can nicely recover the cumulative halo mass functions (CHMFs) when the groups are ranked and binned using a mass proxy. 
Our choice for the mass proxy in this work is obviously the group mass $M_{\rm G}$ of Y21. For each of the 35 lens samples, we define $\hat{M}_{\rm G}$ as its mean mass, and use $n_{\rm G}(> \hat{M}_{\rm G})$ obtained in \S\ref{sec:CHMF_Y21} to denote the halo abundance that the sample corresponds to. We can then use the weak lensing mass, $M_{\rm h}$, of each sample to replace $\hat{M}_{\rm G}$ to form the CHMFs from weak lensing. 
%To obtain unbiased  measurements of the CHMFs in different redshift bins, we need to, (1) calculate the effective group abundance for each of our 35 lens samples, and (2) correct for the halo mass estimation in the CHMFs considering the rank-order change associated with the group mass, $M_{\rm G}$.

%\adb{We first come to obtain the effective group abundance for each of our 35 lens samples. 
%Note that since in each sample, the minimum and maximum group mass $M_{\rm G}$ spans at least 0.2 dex, which results in a large range of maximum and minimum abundances of groups in each sample.  
%We make the following analog to properly obtain the effective group abundance that corresponds to the halo mass we obtained from our ESD measurements. In case the group mass provided in Y21 can trace the true halo mass ideally {\it without any scatter},  then although the group mass, $M_{\rm G}$, may depend on the Planck18 cosmology we assumed,  they would give perfect rank-order of the halo masses of our universe. 
%While from the ESD measurements, the halo mass we obtained in each rank-order bin is the {\it average} halo mass of all the groups in stack, $\hat M_{\rm h}$. Even if the actual cosmology may differ from the Planck18 prediction, the abundance of these halos with average mass $\hat M_{\rm h}$ can be best represented by, 
%\begin{equation}
%    n_{\rm h}(> \hat M_{\rm h})  = n_{\rm G}(> \hat{M}_{\rm G})\,,
%\end{equation}
%where $\hat{M}_{\rm G}$ is the {\it average} group mass in each sample.}

However, as pointed out in D19, 
%neglecting the halo mass scatter induced by the halo mass proxy, e.g., the total luminosity (or the group mass) in the group catalog, the rank-order can be slightly affected. 
the halo mass scatter induced by the mass proxy can lead to a bias in the average halo mass of a mass bin.
%In a particular group mass bin, 
Due to the shape of the halo mass function, there are more contaminations from the lower mass end than those from the higher end, causing the lensing mass to be slightly lower than the ideal case.
%contain {\it more} halos actually from the lower mass bins, and {\it less} halos from higher mass ranges. It will thus result in a halo mass estimated from the lensing observation that is slightly lower than the ideal case.
The cumulative halo mass function obtained from the weak lensing signals is therefore slightly impacted by the mass scatter, especially at the massive end (see Fig. 1 of D19). To correct for such an effect, we use a Monte Carlo method to obtain a mass correction factor, $\Delta \log M$, and define the final lensing mass, $M_{\rm hc}$, with $\log M_{\rm hc} = \log M_{\rm h} + \Delta \log M$. The details regarding the derivation of $\Delta \log M$ are outlined in Appendix \ref{AP:correction}. 

%\adb{After correcting the impact of rank-order change in each of our group mass bin, $\log M_{\rm hc} = \log \hat M_{\rm h} + \Delta \log M$, %[-0.00094062,  0.00156305,  0.00371704 , 0.00879547 , 0.01341748 , 0.02247413,  0.05706843]
%    the final CHMF measurements can be written as,
%\begin{equation}
%    n(> M_{\rm hc})  = n_{\rm G}(> \hat{M}_{\rm G})\,.
%\end{equation}}

We shown in Fig. \ref{fig:chmf1} the CHMFs in the five redshift bins, obtained from our lensing measurement on 35 lens samples in red dots with error bars.
Here since the abundance of the lens systems are rather deterministic, we neglect the error bars along the $y$-axis. 
We provide the values of our measurements in Table \ref{tab:AHMF}. 
%For comparison, we also show in each panel of Fig. \ref{fig:chmf1} using a solid line the theoretical prediction of the cumulative halo mass function \citep{Tinker2008} assuming the Planck18 cosmology. 

\subsection{Discussion}

Compare to the Planck18 model predictions, our observational measurements agree with the model predictions quite well in the intermediate three redshift bins, and are slightly higher and lower in the lowest and highest redshift bins, respectively. 
This discrepancy may be caused by either some possible systematics that our current ESD measurements haven't properly taken into account or the cosmology that we adopted is not the most appropriate one. 
Apart from these, although the weak lensing ESD measurements can provide us direct estimations of the true underlying halo masses, some systematics can also be brought on by effects including asphericity, miscentering, dynamical state, photometric redshift uncertainties, halo member contamination, uncorrelated LSS, assumed density model, and the mass dispersion within the mass bin, etc.  \citep{King_2001,Becker2011,Oguri2011a,Bahe2012,Han2015,Henson_2017,Lee2018,Fong_2019,Grandis_2021}.
%In addition, since we use the NFW profile to describe the halo mass distribution in our lensing model, the weak lensing masses are expected to be biased estimators of the true underlying masses 
%A few systematics of weak lensing mass bias are asphericity, dynamical state, photometric redshift uncertainties, mis-centering, halo member contamination, uncorrelated LSS, assumed density model, and most importantly the mass dispersion within the mass bin can lead to over-estimation or under estimation of the mass.
Here we have taken into account some of the major effects, e.g., photometric redshift uncertainties,
mis-centering and mass dispersion. However, all the other effects may  slightly impact our CHMF results as well. We will come back to this topic by combining weak lensing data from other observations in a forthcoming paper. 

Finally, we note that, as the halo masses we obtained from the weak lensing signals are relatively independent of the cosmology we assumed, except ESD signals somehow rely on $\Omega_{\rm m}$ \citep[see][for related discussions]{More_2013}, we can use the group mass v.s. halo mass relations shown in the upper panels of Fig. \ref{fig:halo_prop} to calibrate/update the group masses in Y21 group catalog. We believe the updated group masses should be less dependent on the assumption of Planck18 cosmology, at least, e.g., the $\sigma_8$ value.

%\adr{XXX Although the deviations of our CHMF measurements are not very significant, it would still be very interesting to investigate in the cosmological parameter space and find which sets can give the best agreement over the whole redshift range. Since in our ESD measurements and modelings, $\Omega_m$ is used, we fix it to the input value. We explore in the  $\sigma_8$ and $m_{\mu}$ parameter spaces for a better model prediction of the overall CHMFs. ...-- in case you are not familiar with such kind of model constraints, you can ask Yu Yu or Liu Yu for help...   XXX} 

\section{conclusions}
\label{conclsion}

In this study, based on the DESI Legacy Imaging Surveys DR9, we probed the dark matter halo properties of 35 lens samples, ranging in redshifts $0.1\leqslant z <0.6$ and group masses $10^{13}$-$10^{15} \msunh$. The 35 lens samples are selected from the group catalogs constructed by Y21 using an extended halo-based group finder. To quantify the halo information, we make use of the weak lensing shear catalogs from the high quality DECaLS data using the Fourier\_Quad method.

We measured the weak lensing signals (ESDs) of these 35 lens samples for three group centroid schemes: the brightest central galaxy (BCG), the luminosity weighted center (LWC) and the number weighted center (NWC), respectively. We take into account the central fraction, off-centering effect and the two halo term contributions in our ESD model to constrain the halo properties for all the lens samples. Our main findings are summarized below:
\begin{itemize}
\item The ESD model can nicely describe the very different subtle features in the observational ESD measurements among different group centroids over a wide scale range. The best-fit halo masses, concentrations and biases are self-consistent among different choices of the centroids. 

\item The off-center component in our ESD model nicely reproduces the off-centering effects at small scales, and can distinguish different center indicators with their very different amounts of center fractions $f_{\rm cen}$.

\item The halo masses we obtained from the ESDs are slightly different from those provided by Y21 based on the abundance matching method discrepancy can be attributed to a small systematic in the group abundance estimation in the lowest redshift bin, systematic errors in the halo mass estimations from the ESD measurements, or the cosmology used to estimate the halo mass. 

\item The concentration - halo mass relations obtained for different redshift bins and halo mass ranges are somewhat higher than the theoretical predictions. %Albeit with large error bars, our observational measurements do not show that halos at higher redshifts have smaller concentrations. 
Unfortunately, the precision of the measurements is insufficient to test the
prediction that halos of a given mass are less concentrated at higher redshifts.
%\adr{Our measurement do not show an upturn after the pivot mass.}

\item The bias - halo mass relations are obtained and also in good agreement with most of the theoretical predictions, except for the lowest redshift bin, in which the observational data prefers sightly lower biases. 

\item By properly taking into account the halo completeness and group purity in the group finding algorithm of Y21, we obtain the number density of groups above the average group mass in each redshift bin. This in turn provides us measurements of the cumulative halo mass functions down to $M_{\rm h}\sim 10^{13} \msunh$ in five redshift bins, that are free from the Eddington bias.

\item Finally, for those who are interested in our data, we have provided the halo properties we extract from our 35 lens samples in Table \ref{tab:pameters}, and the cumulative halo mass function measurements in Table \ref{tab:AHMF}.

\end{itemize}

Our method and results open up new avenues for group/cluster cosmology, provided that stacked weak lensing signals can be accurately measured from high quality imaging surveys and reliable and complete halo systems can be detected from the foreground large (spectroscopic or photometric) redshift galaxy surveys. As we mentioned earlier, these measurements hold important information regarding the cosmology and structure formation information, we will come to this topic in a forthcoming paper.

\begin{acknowledgments}

We sincerely thank the anonymous referee for helpful comments that significantly improved the presentation of this paper.
This work is supported by the National Key Basic Research and Development Program of China (No.2018YFA0404504, 2021YFC2203100), the national science foundation of China (Nos. 11833005, 11890691, 11890692, 11621303, 12073017, 12192224), 111 project No.B20019, and Shanghai Natural Science Foundation, grant No. 19ZR1466800. We acknowledge the science research grants from the China Manned Space Project with Nos. CMS-CSST-2021-A02, CMS-CSST-2021-A01.
 F.Y.D. is supported by a KIAS
Individual Grant PG079001 at Korea Institute for Advanced Study.
The computations in this paper were run on the $\pi$2.0 cluster supported by the Center for High Performance Computing and the Gravity Supercomputer at Shanghai Jiao Tong University.

The Legacy Imaging Surveys of the DESI footprint is supported by the Director, Office of Science, Office of High Energy Physics of the U.S. Department of Energy under Contract No. DE-AC02-05CH1123, by the National Energy Research Scientific Computing Center, a DOE Office of Science User Facility under the same contract; and by the U.S. National Science Foundation, Division of Astronomical Sciences under Contract No. AST-0950945 to NOAO.
The Photometric Redshifts for the Legacy Surveys (PRLS) catalog used in this paper was produced thanks to funding from the U.S. Department of Energy Office of Science, Office of High Energy Physics via grant DE-SC0007914.
\end{acknowledgments}
%~\\
%This paper has been typeset from a \LaTeX file prepared by the author.

\bibliography{main}{}
\bibliographystyle{aasjournal}

\appendix

\section{The average projected density of
the host halo}

According to \cite{Yang2006}, if the candidate lens galaxy (system) is located at the center of host halo, the average projected density of
the host halo can be calculated from the NFW profile, where
\begin{equation}
\rho(r)=\frac{\rho_0}{(r/r_s)(1+r/r_s)^2}\,,  
\end{equation} 
with $\rho_0=\frac{{\bar\rho\Delta_{vir}}}{3I}$, where
$\Delta_{vir}=180$, $I=\frac{1}{c^3}\int_0^c
\frac{xdx}{(1+x)^2}$. Here $c$ is the concentration parameter defined
as the ratio between the virial radius of a halo and its
characteristic scale radius $r_s$, with $r_{\rm 180m}=c\times r_s$.  The projected surface density then
can be analytically expressed as \citep{Yang2006}:
\begin{equation}
\label{eq:SigmaNFW}
\Sigma_{\rm NFW}(R)=\frac{M_{\rm h}}{2\pi r_s^2I} f(x)\,,
\end{equation}
where $M_{\rm h}$ is the halo mass and $f(x)$ bears the following form with
$x=R/r_s$:
\begin{equation}  
f(x)=
\left\{  
  \begin{array}{ll}  
   \frac{1}{x^2-2}[1-\frac{\ln{\frac{1+\sqrt{1-x^2}}{x}}}{\sqrt{1-x^2}}] & x<1  \\  
   \frac{1}{3} & x=1\\  
   \frac{1}{x^2-1}[1-\frac{atan(\sqrt{x^2-1})}{\sqrt{x^2-1}}] & x>1 \,.
  \end{array} 
\right. 
\end{equation}  

On the other hand, if the candidate lens galaxy is not locate at the
center of the host halo, but with an off-center distance $R_{\rm off}$, the
projected surface density will change from an NFW profile
$\Sigma_{\rm NFW}(R)$ to
\begin{equation}\label{eq:Sigmaoff}
\Sigma_{host}(R|R_{\rm
  off}) = 
\frac{1}{2\pi}\int_{0}^{2\pi}\Sigma_{\rm NFW}(\sqrt{R^2+R_{\rm
    off}^2+2R_{\rm off}Rcos\theta}) \, d\theta \,.
\end{equation}

In the halo model, if we consider the mass or galaxy outside the host halo in consideration, which we call as the 2-halo term, we need to take into the halo exclusion effect \cite{Wang2004}. Here we use the following function to describe this effect in the halo-matter cross correlation function,
\begin{equation}\label{eq:exclu}
f_{\rm exc}(r)=
\left\{  
  \begin{array}{ll}  
   0 & r<r_{\rm 180m}  \\  
   1 & {\rm else} \,.
  \end{array} 
\right. 
\end{equation}

\begin{figure*}
    \centering
    \includegraphics[width=1\textwidth]{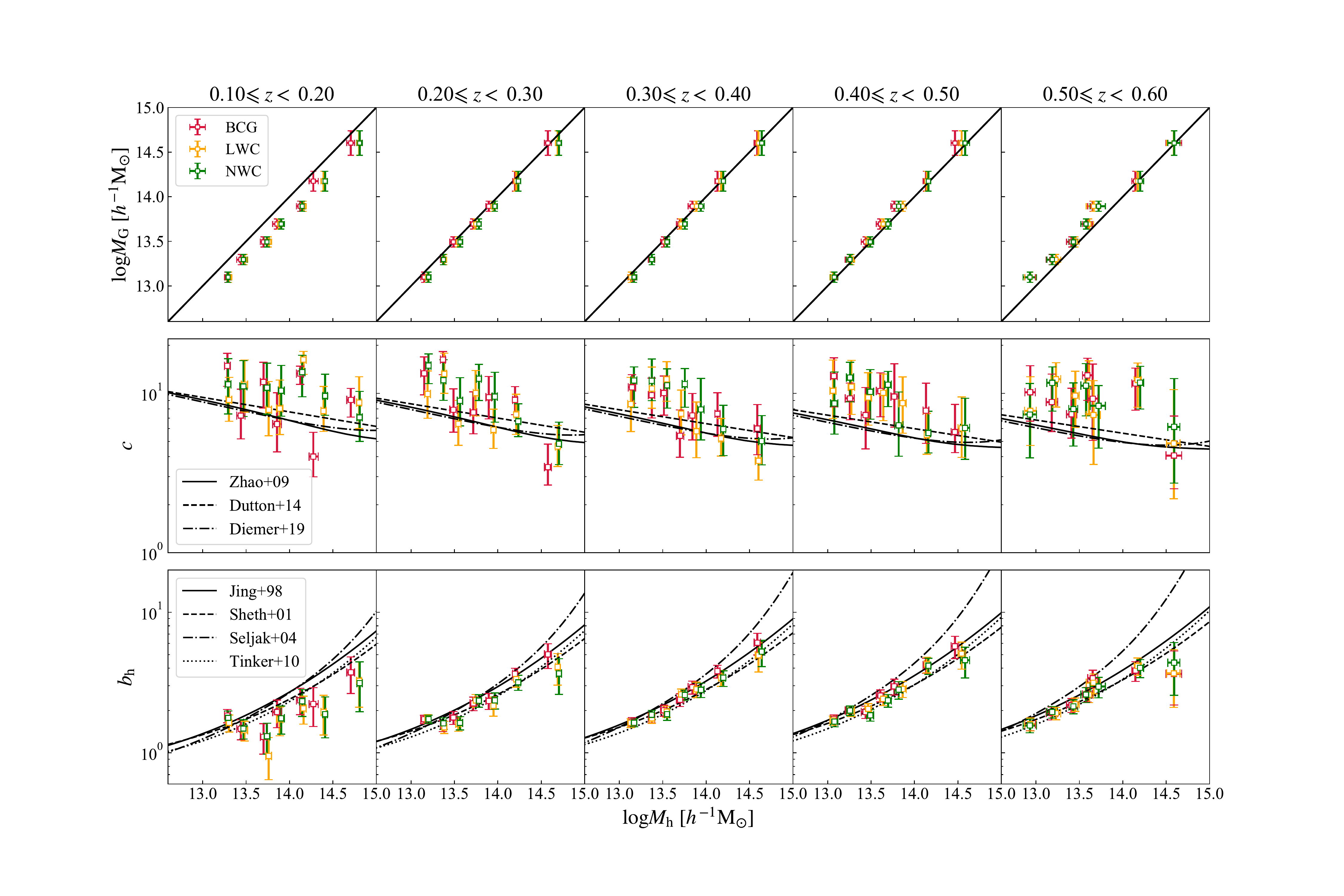}
    \caption{ Similar as Fig. \ref{fig:halo_prop}, but here for the three centroids separately. The red, orange, green points in each panel show the results obtained from signals using BCG, LWC, NWC as center indicators, respectively. 
    }
    \label{fig:halo_prop2}
\end{figure*}

In addition to the halo exclusion effect, to accurately model the cross correlation functions at large scales, one needs to take into account the scale dependence of the halo bias, $\zeta(r,z)$ (see \citet{Fong2022} for a more sophisticated treatment at this scale). Following \citet{Bosch2013}, we start from the one obtained by \citet{Tinker2005}, given by
\begin{equation}\label{zetafit}
\zeta_0(r,z) = {[1 + 1.17 \, \xi_{\rm mm}(r,z)]^{1.49} 
\over [1 + 0.69 \, \xi_{\rm mm}(r,z)]^{2.09} }\,.
\end{equation}
The subscript 0 indicates that this fitting function was calibrated
for halos identified $N$-body simulations using
the friends-of-friends (FOF) percolation algorithm  \citep[e.g.][]{Davis1985}, with a linking length of 0.2 times the mean inter-particle
separation. To take into account the diffidence between this and the one based on the spherical overdensity algorithm,
\citet{Bosch2013} provided a modified version of the bias radial dependence function
\begin{equation}\label{eq:zeta}
\zeta(r,z) = \left\{ \begin{array}{ll}
    \zeta_0(r,z) & \mbox{if $r \geqslant r_{\psi}$} \\
    \zeta_0(r_{\psi},z) & \mbox{if $r < r_{\psi}$}
\end{array}\right.
\end{equation}
where the characteristic radius, $r_{\psi}$, is defined by
\begin{equation}\label{rpsidef}
\log\left[ \zeta_0(r_{\psi},z) \, \xi_{\rm mm}(r_{\psi},z) \right] = \psi\,.
\end{equation}
Here we adopt $\psi=0.9$, the best chosen free parameter obtained in \citet{Bosch2013} for our investigation.

\section{The difference among three halo centering schemes}
\label{AP:THREE}
Shown in Fig. \ref{fig:halo_prop2} are the halo mass, concentration and bias ($\log M_{\rm h}$, $c$ and $b_{\rm h}$) model constraints for our 35 lens samples in different redshift bins as indicated on top of each column separately. The red, orange, green points in each panel show the results obtained for the BCG,  LWC, NWC centering  schemes, respectively.

\section{Mass correction factor associated with the rank-order change}
\label{AP:correction}

As pointed out in D19, a simple way to correct for or reduce the impact of the mass scatter to the measurements of the CHMFs is to add such a scatter into the lensing model, e.g. Eq. \ref{eq:esdmodel}. However, adding such a scatter into Eq. \ref{eq:esdmodel} needs an additional level of integration, which is computationally very time consuming and quite impractical in this study. An alternative and less expensive way is to generate a set of Monte Carlo halos to mimic and correct for the impact of such an effect. We use the following procedures to make our correction:
%We use the following procedures to make our correction. Since there are quite a number of {\it auxiliary} masses will be used in our subsequent presentation, we list in Table~\ref{tab:name} the various masses that are already used in this paper (first four rows) and will be used here (last four rows) along with their definitions in order to avoid confusion.
\begin{enumerate}
    \renewcommand{\labelenumi}{(\theenumi)}
    \item  We first generate a set of halos according to the halo mass function of \citet{Tinker2008} under the Planck18 cosmology,  with true mass $M_{\rm T}$ ranging from $10^{12}$ to $10^{15}\msunh$. We can bin the halos according to the rank-order of their true mass and obtain the average mass $ \left.\hat{M}_{\rm T}\right\vert_{sorted-by-M_{\rm T}}$ inside each bin, which correspond to the corrected halo mass we desired to get.
    %The $M_{\rm T}$ related CHMF is the one we intend to recover.
    \item  Then, we add to each halo a log-normal scatter $\Delta \log M_{\rm T} $ according to the typical halo mass error in the group finder obtained by Y21 using mock data sets. 
    Thus we obtain a scattered mass indicator $\log M_{\rm O} = \log M_{\rm T} + \Delta \log M_{\rm T}$,
    where $\Delta \log M_{\rm T} $ is drawn from a log-normal distribution with a scatter 
    %taken from Fig. 9 in Y21, of which the standard error
    varying from 0.3 dex ($\sim 10^{11.6} \msunh$), 0.4 dex ($\sim 10^{12.3} \msunh$) to 0.2 dex ($\ga 10^{14} \msunh$)(see Fig. 9 of Y21). 
    Here the rank-order of $ M_{\rm O}$ resembles the rank-order of $M_{\rm G}$ in Y21.
    \item  Next, by the rank-order of $M_{\rm O}$, %we can assign each halo with an abundance matching mass $M_{\rm AB}$. This mass resembles the group mass, $M_{\rm G}$, obtained in Y21.
    we can bin the halos and obtain the average halo mass $\left.\hat{M}_{\rm T}\right\vert_{sorted-by-M_{\rm O}}$. In the ideal case where there are no systematic or random errors in the actual lensing measurements, this
    corresponds to the average halo mass $M_{\rm h}$ obtained from the lensing ESD measurement.
    %\item  Within each mass bin, we can get an average mass $\hat{M}_{\rm AB}$, %which again resembles the  average mass $\hat{M}_{\rm G}$ in observation. This mass can be used to obtain the abundance. 
    %\item  , we can obtain the average true halo mass in each rank-order mass bin. 
    %\item  Within this framework, we can obtain the CHMF for halos in each $\hat{M}_{\rm AB}$ mass bin, with average true average halo mass $\hat{M}_{\rm T}$ and abundance $n(\ge \hat{M}_{\rm AB})$. 
    \item  %However, because of , 
    The mass difference between the rank-order by $M_{\rm G}$ and the true halo mass rank-order, which caused by the scatter we introduced, is thus $\Delta \log M = \log \left.\hat{M}_{\rm T}\right\vert_{sorted-by-M_{\rm T}} - \log \left.\hat{M}_{\rm T}\right\vert_{sorted-by-M_{\rm O}}$.
\end{enumerate}

%As the intrinsic CHMF is characterized by halo mass $\left.\hat{M}_{\rm T}\right\vert_{sorted-by-M_{\rm T}}$ and abundance $n(\ge \left.\hat{M}_{\rm T}\right\vert_{sorted-by-M_{\rm T}})$, 
We use $\Delta \log M$ in each bin to correct the lensing mass for the impact of scatter to our CHMF measurements. %based on $\hat{M}_{\rm T}$.
Its value ranges from almost zero in the low mass bins to 0.057 in the most massive bin.

\end{document}